\documentclass[twocolumn]{aastex62}

\newcommand{\bibshift}[1]{}
\graphicspath{{./}{figures/}}

\usepackage{textcomp}
\usepackage{amsmath}
\usepackage{xcolor}
\usepackage{gensymb}
\usepackage[caption=false]{subfig}

\newcommand{\abPan}{0.71435 $\pm$ 0.001}
\newcommand{\abCSP}{0.690 $\pm$ 0.002}

\newcommand{\MBPan}{$-$19.31 $\pm$ 0.04}
\newcommand{\MBCSP}{$-$19.18 $\pm$ 0.03}

\newcommand{\hubblePan}{71.5 $\pm$ 1.8~km/s/Mpc}
\newcommand{\hubbleCSP}{71.5 $\pm$ 1.8~km/s/Mpc}

\shorttitle{Comparing TRGB Distance Scales}
\shortauthors{Anand, Tully, Rizzi, Riess, $\&$ Yuan}

\begin{document}

\title{\textbf{Comparing Tip of the Red Giant Branch Distance Scales: \\ An Independent Reduction of the Carnegie-Chicago Hubble Program \\ and the Value of the Hubble Constant}}

\author{Gagandeep S. Anand}
\affiliation{Space Telescope Science Institute, 3700 San Martin Drive, Baltimore, MD 21218, USA}
\affiliation{Institute for Astronomy, University of Hawaii, 2680 Woodlawn Drive, Honolulu, HI 96822, USA}
\correspondingauthor{Gagandeep Anand}
\email{gsanand@hawaii.edu}

\author{R. Brent Tully}
\affiliation{Institute for Astronomy, University of Hawaii, 2680 Woodlawn Drive, Honolulu, HI 96822, USA}

\author{Luca Rizzi}
\affiliation{W. M. Keck Observatory, 65-1120 Mamalahoa Hwy, Kamuela, HI 96743, USA}

\author{Adam G. Riess}
\affiliation{Space Telescope Science Institute, 3700 San Martin Drive, Baltimore, MD 21218, USA}
\affiliation{Department of Physics and Astronomy, Johns Hopkins University, Baltimore, MD 21218, USA}

\author{Wenlong Yuan}
\affiliation{Department of Physics and Astronomy, Johns Hopkins University, Baltimore, MD 21218, USA}

\begin{abstract}

The tip of the red giant branch has been used to measure distances to 500 nearby galaxies with the Hubble Space Telescope (\textit{HST}) which are available in the Color-Magnitude Diagrams and Tip of the Red Giant Branch (CMDs/TRGB) catalog on the Extragalactic Distance Database (EDD). Our established methods are employed to perform an independent reduction of the targets presented by the Carnegie-Chicago Hubble Program (CCHP) in the series of papers culminating in \cite{2021arXiv210615656F}. Our distinct methodology involves modeling the observed luminosity function of red giant branch and asymptotic giant branch stars, which differs from the edge-detection algorithms employed by the CCHP. We find excellent agreement between distances for 11 hosts with new imaging, all at $D<20$ Mpc. However, we are unable to measure the TRGB for 4 that use archival data designed to measure distances with Cepheids, all at $D>23$ Mpc. With two new \textit{HST} observations taken in the halo of the megamaser host NGC~4258, the first with the same ACS $F606W$ and $F814W$ filters and state of the electronics used for SN Ia hosts, we then calibrate our TRGB distance scale to the geometric megamaser distance. Using our TRGB distances, we find a value of the Hubble Constant of $H_{0}$ = \hubblePan \ when using either the Pantheon or Carnegie Supernova Project (CSP) samples of supernovae. In the future, the James Webb Space Telescope will extend measurements of the TRGB to additional hosts of SN Ia and surface brightness fluctuation measurements for separate paths to $H_{0}$.

\end{abstract}

\section{Introduction} \label{sec:intro}
There is a growing tension in cosmology regarding the value of the Hubble Constant ($H_{0}$). The traditional ``late-universe" approach in measuring $H_{0}$ involves the usage of Cepheid variables to determine distances to nearby type Ia supernova host galaxies all at $z<<1$. The Cepheid distances are used to calibrate type Ia supernova luminosities, that are then applied to SN Ia out into the far-field to measure $H_{0}$. The most recent results of this analysis provide a precise value of $H_{0}$ = 73.0 $\pm$ 1.0 km/s/Mpc \citep{2021arXiv211204510R}. On the other hand, the Planck satellite has collected high resolution data of the cosmic microwave background (CMB). From their data, the Planck team has determined a present value for the Hubble Constant of 67.27 $\pm$ 0.60 km/s/Mpc \citep{2020A&A...641A...6P}. This latter result is typically referred to as the ``early-universe" approach following the naming convention for methods making use of the pre-recombination form of $\Lambda$CDM. The differences in these two methods is nearly a 5$\sigma$ tension for $H_{0}$ at the present day, a fact that has sparked intense discussion about the new physics that would be required in the $\Lambda$CDM model to consolidate these two approaches. This issue is not just apparent in these two studies either-- work by over a dozen separate groups with a wide variety of early and late-universe techniques show similar discrepancies between the two approaches at varying but persistent levels (as recently summarized by \citealt{2021arXiv210301183D}). 

The tip of the red giant branch (TRGB) offers an an alternative tool to Cepheids in the otherwise most-popular rendition of the ``traditional" distance ladder (Geometric Calibrator $\Rightarrow$ Cepheids/TRGB $\Rightarrow$ Type Ia Supernovae). While the TRGB method has become more popular recently due to its role in measuring the Hubble Constant, it has a long history of use and development \citep{1993ApJ...417..553L,2002AJ....124..213M, 2007ApJ...661..815R,2009ApJ...690..389M}. Many groups now routinely use the TRGB as a method of determining distances to nearby galaxies, as it is possible to determine an accurate ($\sim$5$\%$) distance to galaxies within 10~Mpc with a single orbit of \textit{HST} time \citep{2019ApJ...880...52A,2020ApJ...888...31H,2021arXiv210108270B}, with farther distances coming into reach with additional observing time \citep{2017AJ....154...51M,2020ApJ...895L...4D,2021MNRAS.501.3621A,2021ApJ...914L..12S}.

The physical basis for the use of the TRGB as a standard candle lies in the fact that low-mass stars ($<$ 2$M_\odot$) ascend up the red giant branch due to the growth of degenerate helium cores until the cores become hot enough, at $\sim 10^8$~K, to ignite helium. At this instance (known as the helium flash), the star will rearrange its internal structure and present itself on the horizontal branch where it is dimmer by $\sim$4 magnitudes. The key point is that the maximum core mass is almost constant at this event due to a close coupling between core mass and temperature, resulting in the standard candle bolometric luminosity nature of stars at the TRGB. In detail, there are age/mass and metallicity dependencies. More massive stars reach the TRGB at younger ages. Our concern will generally be with red giant branch populations older than 4~Gyr and masses below $1.4~M_{\odot}$.  The more important metallicity dependence largely results from the effects of line-blanketing in the atmospheres of these red giants that causes the release of energy from the star to shift to longer wavelengths with increasing metallicity. The luminosity dependence on metallicity within a specific spectral window is weakest at about the $I$-band, where consequently the technique is generally applied, but can become quite substantial at bluer or redder wavelengths.

\citet{2017A&A...606A..33S} have reviewed the theoretical framework of the evolution of low mass stars in the red giant branch phase of their lives.  As they express, either theoretical models can inform our expectations regarding TRGB luminosities or observations of the TRGB can provide discrimination between models.  Currently the agreement between alternate models and with observations in both absolute luminosities and wavelength dependencies is at the level of $\sim 0.1$~mag.  \citet{2017A&A...606A..33S} consider that a dominant uncertainty with the models lies with bolometric corrections.  Overall, the theoretical models provide confidence in the premise of the TRGB as a tool for measuring distances but the absolute scaling requires an empirical grounding.
For detailed reviews of the method see \citet{2005essp.book.....S},  \cite{2018SSRv..214..113B} and \cite{2021arXiv210402649A}.

\citet{2013AJ....146...86T, 2016AJ....152...50T} used both Cepheid and TRGB distances to set the scale of distances in the Cosmicflows compendium, leading to $H_0$ estimates of $74-76$~km/s/Mpc. \cite{2017ApJ...836...74J} measured the TRGB in the hosts of SNe Ia with archival HST data (most from programs targeting Cepheids) and measured $H_0=73.7$ $\pm$ $2.0$ (stat) $\pm$ $1.9$ (sys) for eight SNe (including two highly reddened SNe Ia, SN 1989B and 1998bu).  Recently, the Carnegie-Chicago Hubble Program (CCHP, \citealt{2016ApJ...832..210B,2019ApJ...882...34F}) targeted eight additional hosts and re-observed three. In a series of papers \citep{2018ApJ...852...60J, 2018ApJ...861..104H, 2018ApJ...866..145H, 2019ApJ...882..150H, 2019ApJ...885..141B,2021arXiv210112232H}, the CCHP team has presented a set of distances to type Ia supernovae hosts from which they ultimately derive their preferred value for $H_{0}$ = 69.8 $\pm$ 0.6 (stat) $\pm$ 1.6 (sys) km/s/Mpc \citep{2021arXiv210615656F}. 

In this paper, we provide an independent analysis of the CCHP data with the well-tested methodology used as part of the Cosmicflows program \citep{2008ApJ...676..184T, 2013AJ....146...86T, 2016AJ....152...50T} and the related Extragalactic Distance Database (EDD, \citealt{2009AJ....138..323T, 2009AJ....138..332J, 2021arXiv210402649A}). Our methodology differs from that used by the CCHP, as we use distinct photometry packages, TRGB measurement techniques, and calibrations for the absolute magnitude of the TRGB. These differences allow us to provide a measurement of the TRGB that is as independent as possible, given that we use the same underlying data.

There has been considerable recent discussion about the absolute calibration of the TRGB (see Table 3 in \citealt{2021ApJ...911...65B} and Section 7.2 in \citealt{2021arXiv210402649A}).  A key ingredient, and one employed by both CCHP and \cite{2017ApJ...836...74J} is a TRGB measurement to the maser host galaxy NGC~4258. Here we consider new \textit{HST} data in two new fields taken in the halo of NGC~4258, the first in this galaxy to be obtained with the same ACS $F606W$ and $F814W$ filters and state of the electronics that have been used to observe the TRGB in the SN Ia hosts. Our calibration is applied to both the Pantheon \citep{2018ApJ...859..101S} and Carnegie Supernova Project/CCHP \citep{2006PASP..118....2H,2017AJ....154..211K} samples of Type Ia supernovae, and we determine values of the Hubble Constant from each.

\section{TRGB Methodology}


\begin{deluxetable*}{lccccccccc}

\tablecaption{Data Summary}
\tablehead{\colhead{Galaxy} & \colhead{Prop. ID} & \colhead{Prop. PI} & \colhead{Blue Filter} & \colhead{Exposure (V)} & \colhead{Exposure (I)} & \colhead{S/N (V)} & \colhead{S/N (I)} & \colhead{$\#$ Stars} & \colhead{$c_{m_{TRGB}}$}}
\startdata
\href{https://edd.ifa.hawaii.edu/get_cmd.php?pgc=34695}{M66}             & 13691        & W. Freedman   & F606W             & 2,420s    & 4,754s    & 2     & 5   & 16,322    & 0.982 \\ \hline
\href{https://edd.ifa.hawaii.edu/get_cmd.php?pgc=32192}{M96}             & 13691        & W. Freedman   & F606W             & 2,410s    & 7,204s    & 2     & 5   & 8,948     & 0.979 \\ \hline
\href{https://edd.ifa.hawaii.edu/get_cmd.php?pgc=50063}{M101}            & 13691        & W. Freedman   & F606W             & 3,650s    & 3,489s    & 5     & 5   & 8,653     & 0.969 \\ \hline
\href{https://edd.ifa.hawaii.edu/get_cmd.php?pgc=12626}{NGC 1309}        & 10497        & A. Riess      & F555W             & 57,600s   & 24,000s   & 2     & 4   & 15,864    & $-$  \\ \hline
\href{https://edd.ifa.hawaii.edu/get_cmd.php?pgc=12651}{NGC 1316}        & 13691        & W. Freedman   & F606W             & 14,676s   & 24,396s   & 2     & 5   & 8,812     & 0.905 \\ \hline
\href{https://edd.ifa.hawaii.edu/get_cmd.php?pgc=13179}{NGC 1365}        & 13691        & W. Freedman   & F606W             & 14,676s   & 24,396s   & 2     & 5   & 9,921     & 0.911 \\ \hline
\href{https://edd.ifa.hawaii.edu/get_cmd.php?pgc=13433}{NGC 1404}        & 15642        & W. Freedman   & F606W             & 37,628s   & 39,946s   & 5     & 5   & 5,724     & 0.713 \\ \hline
\href{https://edd.ifa.hawaii.edu/get_cmd.php?pgc=13727}{NGC 1448}        & 13691        & W. Freedman   & F606W             & 8,551s    & 18,253s   & 2     & 5   & 5,913     & 0.830 \\ \hline
\href{https://edd.ifa.hawaii.edu/get_cmd.php?pgc=28357}{NGC 3021}        & 10497        & A. Riess      & F555W             & 57,600s   & 24,000s   & 2     & 4   & 6,732     & $-$  \\ \hline
\href{https://edd.ifa.hawaii.edu/get_cmd.php?pgc=32207}{NGC 3370}        & 9351         & A. Riess      & F555W             & 57,600s   & 24,000s   & 2     & 4   & 20,682    & $-$  \\ \hline
\href{https://edd.ifa.hawaii.edu/get_cmd.php?pgc=37967}{NGC 4038/9}      & 10580        & I. Saviane    & F606W             & 10,870s   & 8,136s    & 5     & 5   & 40,038    & $-$  \\ \hline
\href{https://edd.ifa.hawaii.edu/get_cmd.php?pgc=39600}{NGC 4258}        & 16198        & A. Riess      & F606W             & 1,059s    & 1,132s    & 2     & 4   & 23,410    & 0.983 \\ \hline
\href{https://edd.ifa.hawaii.edu/get_cmd.php?pgc=40809}{NGC 4424}        & 13691        & W. Freedman   & F606W             & 3,574s    & 10,962s   & 2     & 5   & 15,992    & 0.812 \\ \hline
\href{https://edd.ifa.hawaii.edu/get_cmd.php?pgc=41772}{NGC 4526}        & 13691        & W. Freedman   & F606W             & 3,574s    & 10,962s   & 2     & 5   & 22,854    & 0.849 \\ \hline
\href{https://edd.ifa.hawaii.edu/get_cmd.php?pgc=41823}{NGC 4536}        & 13691        & W. Freedman   & F606W             & 3,536s    & 10,923s   & 2     & 5   & 13,869    & 0.840 \\ \hline
\href{https://edd.ifa.hawaii.edu/get_cmd.php?pgc=51344}{NGC 5584}        & 11570        & A. Riess      & \phantom{*}F555W* & 45,540s   & 14,400s   & 2     & 4   & 13,097    & $-$  \\ \hline
\href{https://edd.ifa.hawaii.edu/get_cmd.php?pgc=51969}{NGC 5643}        & 15642        & W. Freedman   & F606W             & 10,286s   & 10,526s   & 2     & 4   & 13,192    & 0.955 \\ \hline
\enddata
\tablecomments{Table describing the observations used for the comparison within this paper. The individual columns are 1) name of the galaxy, where clicking on the link will open a new browser tab with the respective CMDs/TRGB catalog page, 2) \textit{HST} proposal ID for the underlying data, 3) proposal PI, 4) the bluer filter used to construct the color-magnitude diagram, 5) exposure time in the bluer filter, 6) exposure time in F814W, 7) adopted S/N cutoff in the bluer filter, 8) adopted S/N cutoff in F814W, 9) the number of stars in each galaxy's photometry catalog, and 10) where available, the completeness in F814W at the magnitude of the TRGB. In the table header, we use ``V" and ``I" to denote the approximate \textit{HST} flight filter system counterparts (F555W/F606W and F814W). The ``*" symbol in the ``Blue Filter" column denotes that this observation (of NGC~5584) was taken with WFC3/UVIS instead of the preferred ACS/WFC combination.}
\label{tab:data}
\end{deluxetable*}


An aim of the Cosmicflows team is to independently measure a TRGB distance to \textit{every} galaxy that has the requisite \textit{HST} observations, and to provide this information in an accessible, transparent, and uniform manner through the Extragalactic Distance Database CMDs/TRGB catalog\footnote{\url{edd.ifa.hawaii.edu}}.  To facilitate independent analyses, EDD provides the photometry on which the measurement is based for each host.  Many of these observations have been initiated by the Cosmicflows team itself in order to investigate a variety of topics relating to stellar populations, the constituents of nearby galaxy groups, and large-scale structure. TRGB distances are an important component within the Cosmicflows compilation assembled to study cosmic expansion and deviations from uniform expansion. As of the writing of this paper, the CMDs/TRGB catalog has TRGB distances available for nearly 500 nearby galaxies.

While the details of our methodology have been described in several previous papers \citep{2006AJ....132.2729M,2007ApJ...661..815R,2009AJ....138..332J,2014AJ....148....7W, 2021arXiv210402649A}, an overview is warranted to provide a general sense of our method.  To emphasize the distinctness of our pipeline from the one used by the CCHP team, we highlight three key points:

\begin{enumerate}

    \item We perform PSF photometry and artificial star experiments with DOLPHOT \citep{2000PASP..112.1383D, 2016ascl.soft08013D}. DOLPHOT is well tested and has been used by many groups and projects, including the Panchromatic Hubble Andromeda Treasury (PHAT) survey \citep{2012ApJS..200...18D, 2014ApJS..215....9W}. This software is distinct from the DAOPHOT/ALLFRAME program \citep{1987PASP...99..191S} used by the CCHP group in their main analysis. There is good general agreement between photometry obtained via these two separate packages, although there is the potential for small, systematic offsets in the final photometry at the 0-3$\%$ level \citep{2009ApJS..183...67D, 2010ApJ...720.1225M, 2014ApJS..215....9W, 2021ApJ...906..125J}. We touch upon possible systematics further in Section 3. For this work, we use the recommended parameters presented in the DOLPHOT user's manual\footnote{\url{http://americano.dolphinsim.com/dolphot/}}.
    
    \item Quantifying the position and uncertainty of the TRGB in real data is non-trivial due to the presence of overlapping AGB stars, a finite width of the TRGB, and photometric noise (see Figure \ref{NGC1448} for an example).  The CCHP group measures the location of the TRGB by using an edge-detection algorithm. They first smooth the luminosity function over a selected range of color and magnitude to a chosen smoothing scale and apply a weighted Sobel filter of a specified kernel size and shape to the result. From the output of the edge detection, they identify the midpoint of the dominant peak as the tip of the red giant branch. Our current methodology was first established by \cite{2006AJ....132.2729M} (with influence from \citealt{2002AJ....124..213M}), where they model the luminosity function of asymptotic giant branch (AGB) and red giant branch (RGB) stars as a broken power law, with the break denoting the location of the TRGB. Specifically, the theoretical stellar luminosity function is given the form
    
    \begin{equation}
    \psi = \left\{\begin{matrix}10^{a(m-m_{TRGB})+b}, m-m_{TRGB} \geq 0 \\ 
    \\
    10^{c(m-m_{TRGB})}, \ \ \ m-m_{TRGB} < 0 \end{matrix}\right. 
    \end{equation}

    where $a$ and $c$ are the power law slopes for stars dimmer (RGB stars) and brighter (AGB stars) than $m_{TRGB}$, and $b$ indicates the RGB jump. The artificial star experiments performed within DOLPHOT are used to quantify the levels of error, bias, and completeness present in the measured photometry, and these are all explicitly taken into account when modeling the observed luminosity function of AGB and RGB stars. Further work by \cite{2014AJ....148....7W} present further refinements to improve on speed. The underlying methodology mentioned here has formed the basis for a large amount of work on the TRGB done by several groups (e.g. \citealt{2011AJ....141..106J,2014MNRAS.443.1281K, 2017AJ....154...51M, 2018MNRAS.474.3221M, 2019ApJ...872...80C}, among many others). Full details of our methodology are available within \cite{2006AJ....132.2729M}, \cite{2014AJ....148....7W}, and \cite{2021arXiv210402649A}, the latter of which provides a decadal update on the features (e.g. color images) of the CMDs/TRGB catalog first presented by \cite{2009AJ....138..332J}.
    
    \item Unlike the CCHP who use a singular value for $M_{TRGB}$, we standardize the TRGB measurement to a fiducial color following \cite{2007ApJ...661..815R} to include a color term to account for the modest metallicity dependence of the location of the TRGB in the $F814W$ filter. While the CCHP use the bluest red giant branch stars for their analysis of each galaxy, these stars do not necessarily have the same color distribution from galaxy to galaxy. The zero-point of the \cite{2007ApJ...661..815R} calibration used in the EDD database was set via horizontal branch measurements in five nearby dwarf spheroidal galaxies, ultimately linked to trigonometric observations \citep{2000ApJ...533..215C}. This calibration is entirely independent from the calibration in the LMC used by the CCHP, and is given by, for the F606W and F814W filters:
    
    \begin{multline}
        \mu_{TRGB} = (m_{TRGB}-A_{F814W}) - \\ (-4.06 + 0.2[(F606W-F814W)_{0} - 1.23]
    \end{multline}
    
    where $A_{F814W}$ is the adopted foreground extinction and $(F606W-F814W)_{0}$ is the median color of the TRGB, after correcting for reddening. In any event, in this paper, we will use the maser distance to NGC~4258 to calibrate the TRGB and reference both the CCHP and our measurements to this common zero point peg.

\end{enumerate}

Though not intentionally (as the established procedures for the CMDs/TRGB catalog go back over a decade), we find that our data reduction and analysis techniques provide for TRGB distance measurements that are as independent from the CCHP team as we think is possible, given the underlying imaging is the same.

\section{Data}

We retrieved the individual \textit{*.flc} \textit{HST} images from the Mikulski Archive for Space Telescopes (MAST\footnote{\url{https://archive.stsci.edu/hst/search_retrieve.html}}). These images are already corrected for issues that can arise due to charge transfer inefficiencies present with the \textit{HST} cameras. For purposes of alignment within DOLPHOT, we create drizzled (\textit{*.drc}) images using \textit{DrizzlePac} \citep{2015ASPC..495..281A}. We stress that we do not use the drizzled images for photometry, as the resampling done to create these drizzled images can affect the accuracy of the final measurements. The photometry is performed directly on the individual \textit{*.flc} exposures (and subsequently combined by DOLPHOT), with the drizzled image serving as the basis for object detection as well as a coordinate reference frame\footnote{See \href{https://physics.mcmaster.ca/Fac_Harris/dolphot_primer.txt}{this} unofficial DOLPHOT guide for more details.}. A summary of the observations used in this paper is shown in Table \ref{tab:data}.

Once the combined photometry and artificial stars are complete, we apply quality cuts to both using a modified version of the criteria presented in \cite{2017AJ....154...51M}. Namely, we select for stars which meet the following criteria:

\begin{itemize}
    \item Object Type $\le$ 2
    
    \item Error Flag $=$ 0 in both bands

    \item $(Crowd_{\mathrm{F555W/F606W}}+Crowd_{\mathrm{F814W}}) < 0.8$
    
    \item $(Sharp_{\mathrm{F555W/F606W}}+Sharp_{\mathrm{F814W}})^{2} < 0.075$
\end{itemize}

To describe these selections, we briefly summarize the information presented in the DOLPHOT user's manual. Object types of 1 or 2 select for only ``good" or ``faint" stars, respectively, and stars with an error flag = 0 includes only stars which were determined to have been well-recovered. The sharpness criteria (zero for a perfect star) allows us to exclude cosmic rays (large positive sharpness values) and background galaxies or blends of stars (large negative sharpness values). Lastly, use of the crowding parameter allows us to remove stars which lie in highly crowded environments. The only metric we vary from galaxy to galaxy is the adopted signal-to-noise cutoff, as the relative depth of the imaging varies for each target. We provide the final signal to noise selections used for each target in Table \ref{tab:data}.  

We note that we reduced this dataset as independently as possible from the CCHP team. Our database has hosted the present TRGB distance to NGC~4038/9 since 2010. For the remainder of the targets under consideration, we performed reductions and TRGB analyses as data became publically available from their respective programs on the \textit{HST} archives (and as time allowed), and subsequently made the results publicly available on EDD. As a consequence, distances to several of the CCHP selected galaxies (M66, M96, M101, NGC~4038/9, $\&$ NGC~5643) were first available within our catalog, whereas the rest were first published by the CCHP team. 

Additionally, an important note about the photometric uncertainties determined throughout this work (and many others in the literature)$-$ they are very likely to be underestimates, and in a way which is difficult, if not impossible to quantify with the present dataset. Besides differences in choices of PSF fitting software and their respective suite of options, extensive work by the ANGST and PHAT collaborations \citep{2009ApJS..183...67D, 2012ApJS..200...18D, 2014ApJS..215....9W} have uncovered systematic offsets ($\sim$0.02$-$0.05 mag) in PSF magnitude measurements of identical stars in different locations on \textit{HST's} detectors. These systematic offsets are not only dependent on position, but also magnitude. The source of these systematic offsets likely lie in some combination of the adopted CTE-corrections, flat-fielding methods, and PSF libraries used. While the ACS and WFC3 instrument teams are continuously improving their methods (e.g. \citealt{2017acs..rept....2H, 2018acs..rept....4A, 2018wfc..rept...14A}), some fraction of these issues likely persist. Without the wealth of data that these previous large surveys have covered, we are simply unable to quantify the analogous uncertainties in our data. Instead, we simply advise the reader to keep this additional, currently unquantified source of uncertainty ($\sim$0-3$\%$ in distance) in mind.

\begin{figure*}
\epsscale{1.15}
\plotone{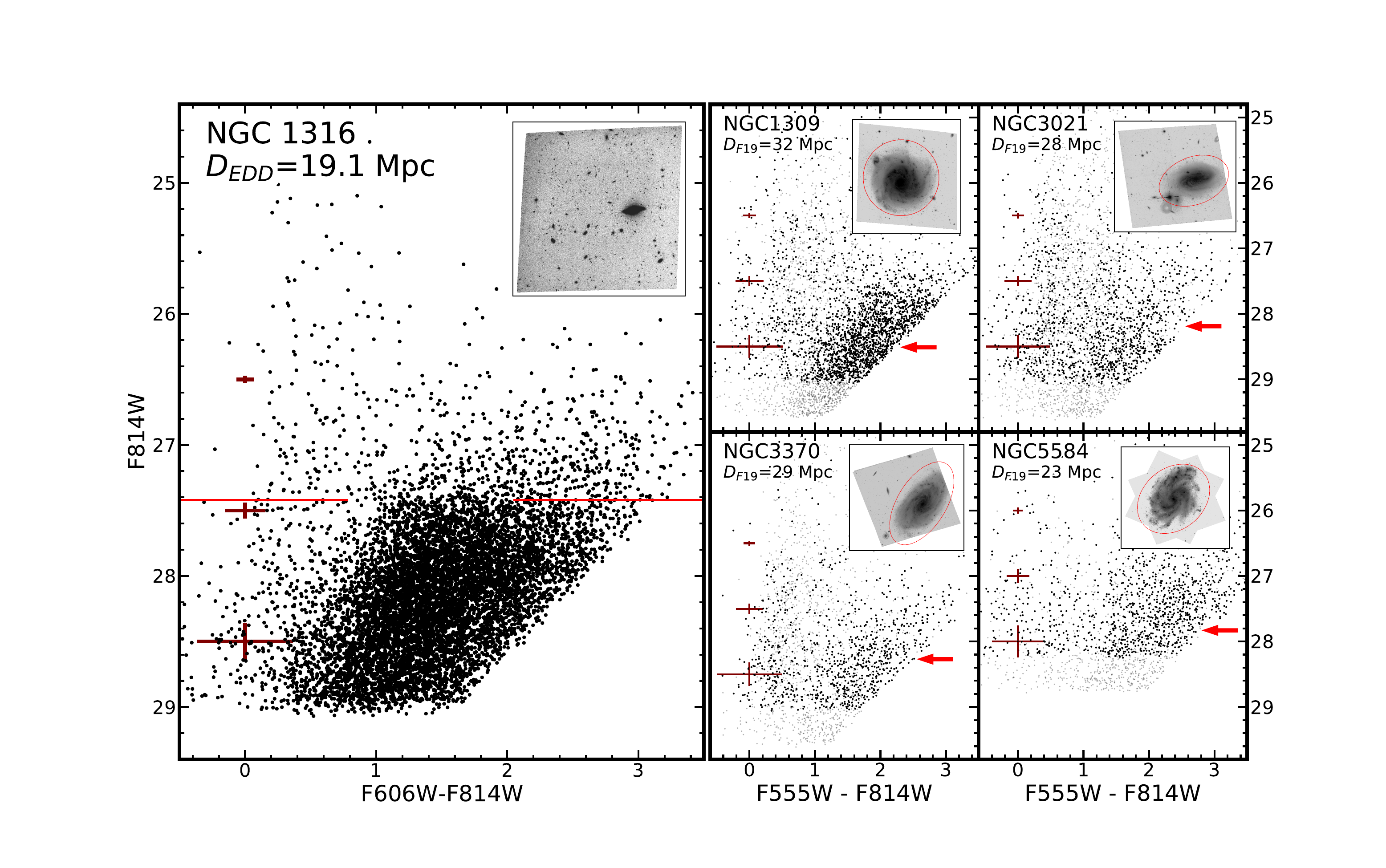}
9\caption{\textbf{Left)} CMD of NGC~1316, for which we derive a TRGB distance of 19.1 $\pm$ 0.9~Mpc. The HST \textit{F814W} field of view is shown in the inset. The gap in the broken red line shows the color range used for our determination of the TRGB. Magnitude and color error bars (calculated at the measured color of the TRGB) as reported by DOLPHOT are shown on the left in maroon. \textbf{Right)} Same as the left panel, but for the four Cepheid hosts for which we are unable to derive TRGB distances. In black, we use the same photometric cuts for these CMDs as we do for the rest of the sample, and apply the least-stringent S/N selections used for any galaxy in this study (2 in F555W, and 4 in F814W). These CMDs are limited to the regions exterior to the red ellipses shown in the \textit{HST} $F814W$ insets. The values of $m_{TRGB}$ as implied by the distances reported in \cite{2019ApJ...882...34F} are shown as red arrows on each diagram. The representative color errors are calculated at an approximate TRGB color of F555W-F814W = 2.0. None of these CMDs show a clear discontinuity corresponding to the TRGB in the expected regions or any others, and are all shallow when compared to the new CCHP data. In gray, we show the same data, but with less stringent quality cuts (S/N = 2.5 in F814W,  weaker cuts on sharpness, and no crowding cuts). }
\label{CepheidHosts}
\end{figure*}

\subsection{CCHP Sample Quality}

As the CCHP team gave careful consideration to the placement of their \textit{HST} fields, we see that this data is similar in nature to the most reliable fields for other galaxies which exist on the CMDs/TRGB catalog. Specifically, the fields are well-placed in the halos of these supernova host galaxies, with minimal contamination from younger stellar populations. In cases where there are sizeable regions of visible young stars (e.g. NGC~1448), we exclude these regions from our analysis. The one additional exception is NGC~1404, for which the observations are taken closer to the core of the galaxy. For this reason, we elect to remove more than half the detector from our sample due to higher than typical levels of crowding.

\subsection{Archival Sample Quality}

While we find that the new dataset obtained by the CCHP team is well-suited for measuring the TRGB, the same cannot be said for the set of older, archival data used in their analysis. Most of this data (with the exception of NGC~4038/9) was obtained to detect brighter Cepheids, and the observing programs targeted the disks and thus their fields of view include only a small fraction of the halos of these galaxies.

Take for comparison the CMD of NGC~1316 shown in the left-hand panel of Figure \ref{CepheidHosts}, which we will find to lie at a distance of 19.1~Mpc. The data for this galaxy was obtained by the CCHP (GO-13691, PI W. Freedman). In $F814W$ (the filter used for measuring $m_{TRGB}$), the total exposure time is 24,396s. The ``bluer" filter for this observation, which is used to isolate the red giant branch from younger stellar populations, is the preferred $F606W$. $F555W$ can be used to construct a CMD and measure the TRGB but $F606W$ is preferred because 1) the lesser infrared sensitivity of $F555W$ causes the redward clip on the CMD to impinge substantially on the red giant branch, and 2) $F606W$ is a significantly wider filter than $F555W$\footnote{See the ACS/WFC throughput curves at \url{https://www.stsci.edu/hst/instrumentation/acs/data-analysis/system-throughputs}.}, allowing more flux to reach the detector (there is an increase in depth of 0.7 mag for equal exposure times).

Now give attention to the archival sample: NGC~1309, NGC~3021, and NGC~3370 all have 24,000s of observations in $F814W$, essentially the same as NGC~1316. The right-hand panel of Figure \ref{CepheidHosts} shows the CMDs for each of these galaxies, trimmed to exclude regions that lie within the crowded and dusty disks (seen in the inset imaging). In displaying these main CMDs, we use the same quality cuts as described earlier, and adopt the least stringent S/N selections as we do for any other galaxy in this study (S/N = 2 in $F555W$, and S/N = 4 in $F814W$). Our photometry for these targets reach a similar depth as it does for NGC~1316 ($F814W\sim$29 mag). The expected values of $m_{TRGB}$ are nearly 1 magnitude fainter for these galaxies than is the case with NGC~1316. A crude estimate of the exposure times required to reach the same S/N ratio at $m_{TRGB}$ is $2.5^{2} \sim$6$\times$. A second set of data is also plotted, in a dimmer grey color. For this sample, we lowered the S/N cut in F814W to 2.5, removed the crowding cuts, and weakened the (relatively stringent) sharpness cuts to $\le |0.3|$ in both bands (the base recommendation from the DOLPHOT manual).

Beyond the limitation in depth, these observations have additional shortcomings. The placement of these archival fields in the disks (originally targeted for Cepheids) increases the potential of errors from crowding. Additionally, the bluer filter used for this set of data is $F555W$, instead of the preferred $F606W$, which causes part of the RGB to be ``pushed" off the right edge of the observed CMD beyond the credible detection limit. While partially mitigated by the increased exposure time in $F555W$ relative to F814W, the observations are targeting the disks of these galaxies, which are inherently more metal rich (and hence redder and fainter in $F555W$). And while observations for a fourth supernova host (NGC~5584) are taken with what seemingly should be sufficient depth, the data for this target were taken with WFC3/UVIS instead of the standard ACS/WFC. The overall system throughput for WFC3 is decreased by nearly a factor 2 in $F814W$ (see Figure 2 in \citealt{2018wfc..rept....2D}). In addition, the disk of NGC~5584 covers most of the smaller WFC3/UVIS field, leaving only $\sim$ 15\% of the field useful for detecting halo stars.

As noted earlier, one exception to the above issues involves the Antennae galaxies (NGC~4038/9). The archival observations for this target (GO-10580, PI I. Saviane) exhibit a confluence of preferred qualities. Namely, the camera and filters (ACS/WFC with $F606W$+$F814W$) are optimal for a measurement of the TRGB, and the field placement is in a less crowded region of the galaxy (away from the merging disks and on an outer tidal tail). This circumstance is not an accident, as these archival observations were taken with the primary goal of measuring the distance to the Antennae via the TRGB \citep{2008ApJ...678..179S, 2008AJ....136.1482S}. 

\begin{figure*}
\epsscale{1.1}
\plotone{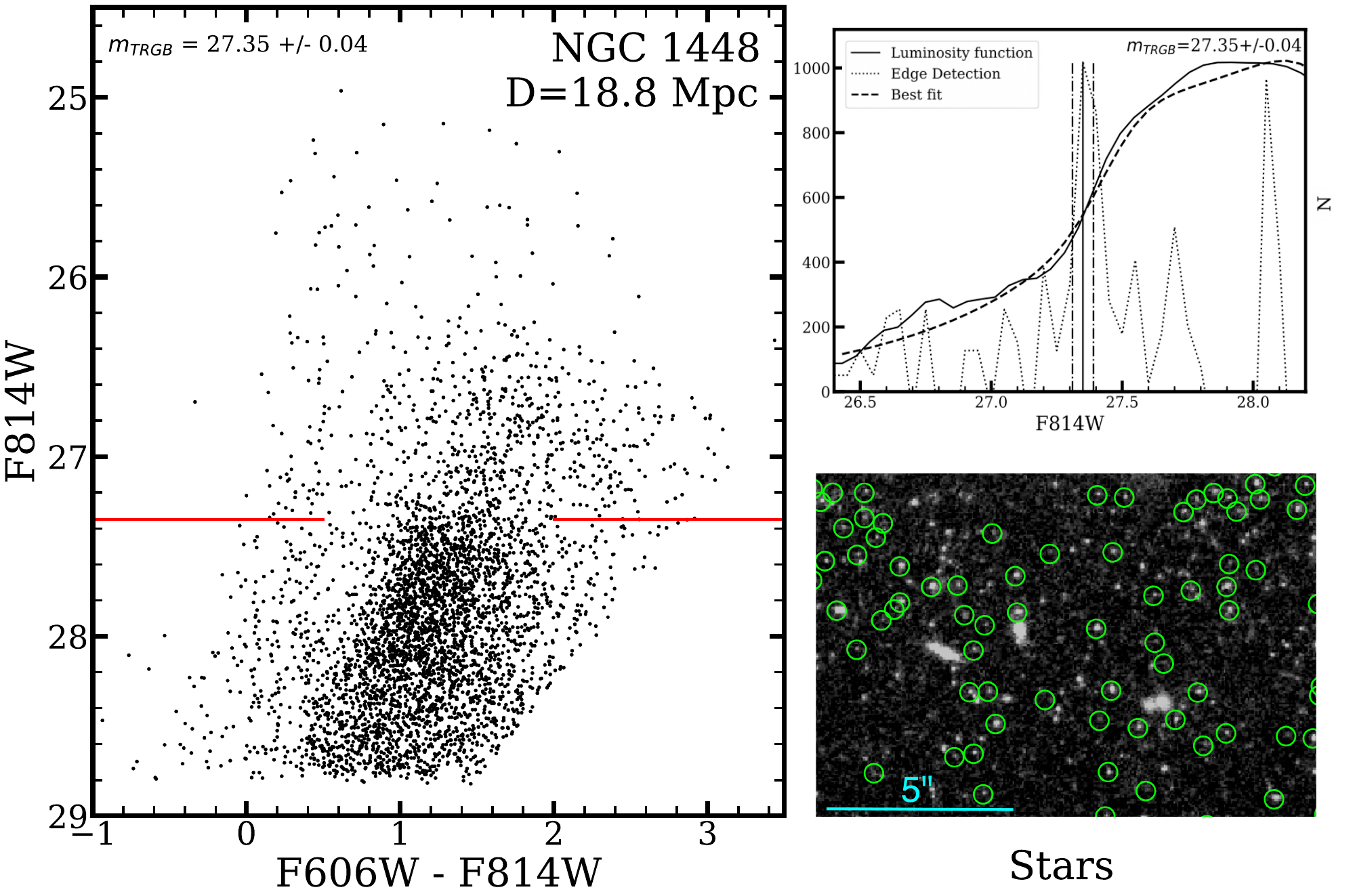}
\caption{\textbf{Left)} Color-magnitude diagram for NGC~1448, which we find to lie at 18.8 $\pm$ 0.8~Mpc. The gap in the broken red line shows the color range used for our determination of the TRGB. \textbf{Top Right)} The observed luminosity function (solid line), our model best-fit luminosity function (dashed line), and results of a first-derivative edge-detection algorithm (dotted line, shown for visual comparison). \textbf{Bottom Right)} A cutout of the F814W drizzled reference frame, with stellar detections that pass our quality cuts overplotted as green circles. This region is near the top of the detector (see Figure \ref{starCollage} for a look at the entire detector).}
\label{NGC1448}
\end{figure*}

\section{TRGB Distances}

In this section, we focus on a comparison between the CCHP and EDD TRGB results derived from the same underlying data. Afterwards, we provide a measurement of the TRGB in the megamaser host NGC~4258 based on new \textit{HST} data obtained at two positions in the halo of the galaxy. These combined results allow a comparison of our two TRGB distance scales.

All our photometry and TRGB distance information, as well as footprints, CMDs, and color images are publicly available within the CMDs/TRGB catalog on EDD (hyperlinks to each galaxy's page can be found in the first column of Table \ref{tab:data}). A complete description of the contents of the catalog is provided by \cite{2021arXiv210402649A}. We also provide our determination of the observed magnitude of the TRGB for each galaxy ($m_{TRGB}$), as well as the measured median colors at the tip in Table~\ref{tab:H0}. Additionally, we provide footprints, plots of stars used for developing our CMDs, the CMDs themselves (along with TRGB measurements), and luminosity functions for our reductions of the CCHP targeted galaxies in Appendix A (see Figures \ref{footprintCollage}, \ref{starCollage}, \ref{CMDcollage}, and \ref{fig:LFcollage}).

\subsection{CCHP Sample Results}

We are able to reliably detect the TRGB for the sample of 11 galaxies with new data obtained by the CCHP group from GO-13691 and GO-15642 (PI. W. Freedman). An example color-magnitude diagram from our reduction of this dataset is shown in Figure \ref{NGC1448}, with information on the remaining galaxies shown in Appendix A (Figures \ref{footprintCollage} $-$ \ref{fig:LFcollage}). The left panel shows the CMD and TRGB determination for NGC~1448, for which we derive a distance of D=18.8 $\pm$ 0.8~Mpc. The observed luminosity function, best-fit model luminosity function, and results of a first derivative edge-detection (displayed for visual comparison) are shown in the top-right hand panel. The bottom-right hand panel shows a small portion of the drizzled F814W image used as the reference frame, with stellar detections that pass the DOLPHOT quality cuts encircled in green. The region shown is close to the top of the detector, where crowding is most significant, yet the stellar detections are still cleanly separated. Portions of the very top of the chip had been manually excluded from our analysis due to visible star formation, which had been measured via the presence of large amounts of blue, upper main sequence stars (see Figure \ref{starCollage} for a look at the full detector). 

\subsection{Archival Sample Results}

We do not have great success with measuring the TRGB in the sample of galaxies with archival data. In fact, we are only able to obtain a TRGB measurement for one (the closest and which did not target the disk) out of the five galaxies that fall into this category. We find the value of $m_{TRGB}$ for the Antennae galaxies (NGC 4038/9) with modest statistical precision ($\sim6\%$), with non-detections for NGC~1309, NGC~3021, NGC~3370, and NGC~5584, all expected to be at $D>23$ Mpc. Given our discussion of the underlying data quality in Section 3.2, this is not unexpected. These four galaxies lie \textit{significantly} further away than the rest of the sample, with a measured distance range of 28$-$32~Mpc based on their SNe \citep{2019ApJ...882...34F}. 
The distance modulus in the CMDs/TRGB catalog for the fifth galaxy, NGC~4038/9 (for which the observations were designed to measure a TRGB distance, and is closer than the above four targets) is $\mu$ = 31.72 $\pm$ 0.14 (D = 22.1 $\pm$  1.43 Mpc), similar to the value of $\mu$ = 31.68 $\pm$ 0.06 measured by \cite{2019ApJ...882...34F}.

In the literature, TRGB distances to the four distant Cepheid hosts was first presented by \cite{2015ApJ...807..133J} and \cite{2017ApJ...836...74J}. In the latter work, they also presented a measurement of the Hubble constant of $H_{0}$ = 71.2 $\pm$ 2.5 km/s/Mpc. We note that the photometry presented in these papers was performed directly on drizzled images, which is generally not recommended as the resampling of the PSF can lead to small but non-zero systematic offsets in the final photometry. The photometry in these works probes deeper than what we find with our analysis, which is likely a result of differences in adopted quality cuts on the initial photometric catalogs.

In \cite{2019ApJ...882...34F}, the CCHP presented a compilation of TRGB distances, as well as their first measurement of the Hubble Constant. In this work, they remeasured the distances to these hosts with archival data, and found good agreement with the exception of NGC~5584, for which they increased the reported uncertainty to reflect this situation. In their distance compilation, they adopt the TRGB measurements as found by \cite{2017ApJ...836...74J}, although they update the color-dependent zero-point from the one presented by \cite{2017ApJ...835...28J} to their preferred flat value. Due to the low signal to noise ratios combined with the incomplete coverage of the red giant branch in the CMD (due to the usage of the F555W filter), we are not able to perform measurements of the TRGB with our photometry for these galaxies with our methodology.

Our failure to measure distances to four of the supernova hosts (NGC~1309, NGC~3021, NGC~3370, and NGC~5584) is not surprising given the inferior relative depths compared to the anticipated TRGB magnitudes and the field placements that result in higher levels of crowding, population confusion, and internal extinction. We note that \cite{2019ApJ...882...34F} found excellent agreement between all of the Cepheid and TRGB distances to the hosts targeted by the CCHP but poor agreement for three of the five cases measured with archival data. The difficulty in measuring the TRGB in these archival data sets may play a role in two of the three mismatches (NGC~3021 and NGC~3370, but not NGC~4038/9).  We do not believe these galaxies should be currently considered for the important task of calibrating supernova distances as the quality of the underlying data are simply not comparable to that of the newer CCHP observations. While being unable to measure TRGB distances to these galaxies, we still provide our color-magnitude diagrams and underlying DOLPHOT photometry to these targets via our CMDs/TRGB catalog.

\subsection{NGC~4258}

\begin{figure*}
\epsscale{1.03}
\plotone{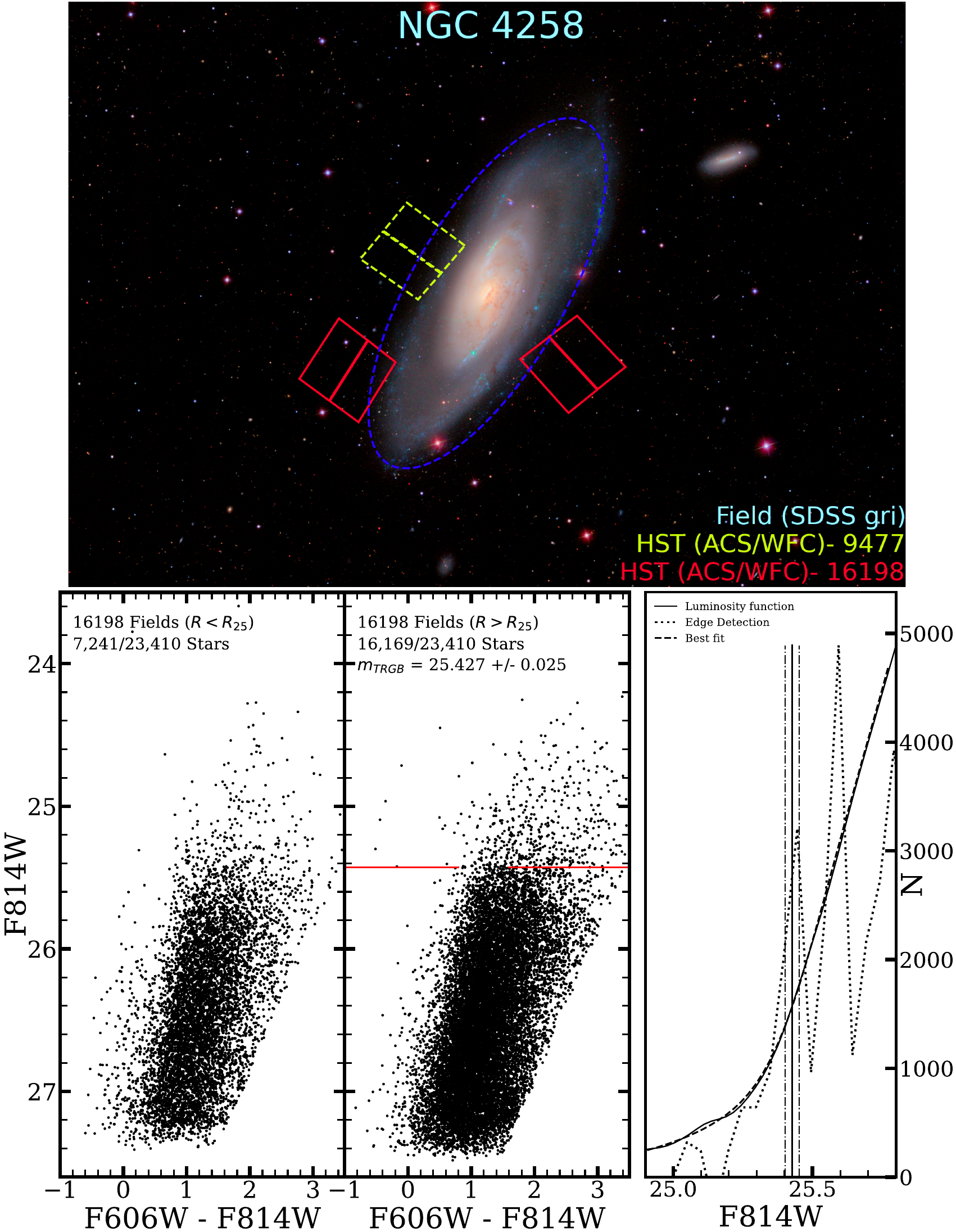}
\caption{\textbf{Top)} A 30'$\times$40' SDSS \textit{gri} color composite of NGC~4258. North is up and east is to the left. The yellow dashed footprint shows the field used for the previous CMDs/TRGB catalog determination of $m_{TRGB}$. The new fields from GO-16198 (PI A. Riess) are shown in red. We use the regions of these fields exterior to $D_{25}$ (shown in dashed blue) for our final measurement of $m_{TRGB}$. \textbf{Bottom Left)} CMD for the combined regions of GO-16198 interior to $D_{25}$. All of these stars are excluded from our analysis, due to the possibility of small amounts of dust extinction from within the spiral disk. \textbf{Bottom Middle)} CMD for the combined fields from GO-16198, excluding regions which fall within $R_{25}$. Our final measurement of the TRGB is $m_{TRGB}$ = 25.427 $\pm$ 0.025. The gap in the broken red line shows the color range used for our determination of the TRGB. \textbf{Bottom Right)} Our model fit to the luminosity function for our final TRGB determination of NGC~4258. Results of a first-derivative edge detection algorithm are shown for visual comparison.}
\label{NGC4258}
\end{figure*}

NGC~4258 has a very precisely known distance of D~=~7.568 $\pm$ 0.082 (stat) $\pm$ 0.076 (sys)~Mpc obtained from observations of its water megamaser \citep{2019ApJ...886L..27R}. The previous CMDs/TRGB catalog distance determination for NGC~4258 was $7.65\pm0.12$~Mpc based on a TRGB measurement of $m_{TRGB}$ = 25.43 $\pm$ 0.03 at a median color of F555W-F814W = 2.12 $\pm$ 0.04. This determination was derived from the outer regions ($\sim$25$\%$) of the field shown in dashed yellow of Figure 3 (GO-9477, PI. B. Madore). 

As part of a Cycle 28 \textit{HST} program (GO-16198) designed to increase the sample of Cepheids in NGC 4258, A. Riess and collaborators obtained two fields of parallel ACS $F606W$ and $F814W$ imaging in the halo of NGC~4258. The footprints for these two ACS fields are shown in red in the top panel of Figure \ref{NGC4258}, where the background image is a composite color (\textit{gri}) image created from Sloan Digital Sky Survey imaging \citep{2000AJ....120.1579Y}. 

While great effort has been put into measuring the magnitude of the TRGB in this galaxy \citep{2006ApJ...652.1133M, 2007ApJ...661..815R, 2008ApJ...689..721M, 2021ApJ...906..125J}, our new observations have two key benefits over all previous fields used in literature measurements. First, the bluer filter for these new fields is the preferred $F606W$ instead of F555W. All the other TRGB measurements presented in this paper use $F606W$ as the bluer filter, which allows for a more accurate comparison. Second, these are the first fields taken in the halo of NGC~4258 after the May 2009 mission (STS-125) that serviced the electronics and replaced the power supply on ACS and with the same or similar state of charge transfer efficiency degradation. There is a 10-15 year gap between 2003-2005 when the ACS observations of NGC 4258 used to calibrate TRGB by \citet{2021ApJ...906..125J} were obtained and those in the 2015-2020 time period of the SN host observations and the new NGC 4258 data.  In this span a TRGB star in the halo of NGC 4258 is estimated to suffer a change of 0.1 mag due to degrading CTE. While this is largely corrected by pixel-based CTE rectification in the pipeline, the correction is a model which is imperfect and it is likely some fixed fraction of the correction persists as an error in $H_0$ which is best reduced by using data for NGC 4258 and the SN Ia hosts for which the electronics are in a more similar state.

Our initial TRGB measurement was $m_{TRGB}$ = 25.444 $\pm$ 0.031, from a combination of both of our new ACS fields. However, as shown by the recent careful analysis of \cite{2021ApJ...906..125J}, it is possible for contamination from the outer disk of the galaxy to introduce small systematic deviations in the measured magnitude of the TRGB. It is not clear that our distinct methodology would suffer from these same issues, however we opt to be conservative in our analysis. To avoid any contamination from the outer disk of NGC~4258, we further restrict the selection of stars used for the final measurement by removing any stars that lie within $D_{25}$, which corresponds to a semi-major axis SMA (semi-major axis) $>$ 10'. This radius (shown as the dashed blue line in the top panel of Figure \ref{NGC4258}) is defined at a level of 25 $\mathrm{mag/arcsec.^{2}}$ in B-band, as reported by \cite{1991rc3..book.....D}. The isophotal radius of the sources in our two fields as seen in Figure \ref{NGC4258} thus range between 10' to 16' SMA, similar to the range of radii used by \cite{2021ApJ...906..125J}. This selection is also at a greater isophotal radius than the GO-9477 field (though similar to the region used for the previous CMDs/TRGB catalog analysis of this archival field).

Using this radius to trim our sample reduces the number of stars in the final CMD from 23,410 to 16,169. The removed stars are shown in the bottom-left hand panel of Figure \ref{NGC4258}. Our final result is shown in the bottom-middle panel of Figure \ref{NGC4258}, where we find $m_{TRGB}$ = 25.427 $\pm$ 0.025 at a median color of $F606W-F814W$ = 1.33 $\pm$ 0.04. The bottom-right panel shows the luminosity function and our best-fit result. The results of a first-derivative edge detection are also shown for comparison and match well with our modeled break. Our final result is fainter than the recent measurement of $m_{TRGB}$ = 25.372 $\pm$ 0.014 by \cite{2021ApJ...906..125J} by 0.055 $\pm$ 0.028 mag, though the results are consistent at a level of 1.9$\sigma$ given the mutual statistical uncertainties added in quadrature. As discussed in Section 3, the significance of this disagreement is likely an overestimate, as there are various small-scale systematics which are not easily brought out with such limited datasets. Our new measurement is nearly the same as the previous EDD determination from the outer 25$\%$ of the GO-9477 field ($m_{TRGB}$ = 25.43 $\pm$ 0.03). 

We can now compare TRGB distances to NGC~4258. In our case (EDD), $\mu_{N4258}^{EDD} = m_{TRGB} -A_{F814W} -M_{TRGB} = 25.427-0.025+4.044 = 29.446 \pm 0.079$ mag. In the case of CCHP, $\mu_{N4258}^{CCHP} = 25.372 - 0.025 + 4.045 = 29.392 \pm 0.039$ mag, where the adopted foreground extinction $A_{F814W}$ is the same and the value of $M_{TRGB}$ is adopted from \citet{2021arXiv210613337H}. The corresponding distances are $D_{N4258}^{EDD} = 7.748~\pm~0.282$~Mpc and $D_{N4258}^{CCHP}=7.558~\pm~0.136$~Mpc.

\subsection{EDD and CCHP Comparison}

\begin{figure*}
\epsscale{1.15}
\plotone{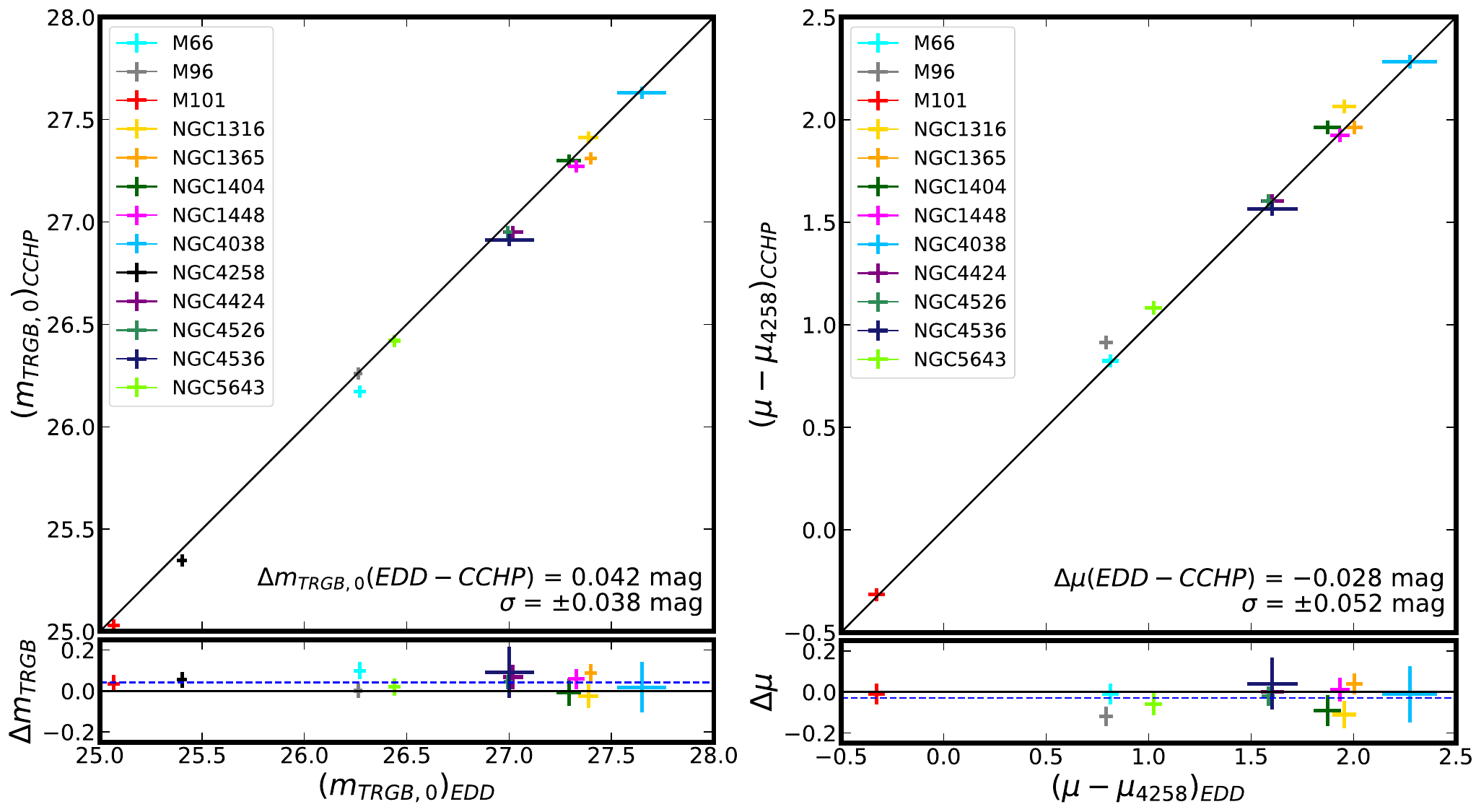}
\caption{\textbf{Left)} A comparison between the measured value of the TRGB $m_{TRGB,0}$ by CCHP and ourselves (EDD). Overall, the EDD catalog reports values that are 0.042 mag fainter (dashed-blue line), averaged by host, than those found by the CCHP. EDD values typically probe further red in color than CCHP. Since the $F814W$ magnitude of the TRGB is known to decrease slightly with redder color, this small offset is expected. \textbf{Right)} A comparison of the CCHP and EDD catalog TRGB distance scales relative to the maser host NGC~4258 distance.  Overall we find good agreement between our two distance scales, bolstering confidence in the ability of the TRGB to provide high-quality, reproducible distance measurements. The dashed-blue line in the bottom panel shows the mean offset ($-$0.028 mag) weighted by galaxy between our two \textit{relative} distance measures (subtracting each group's value of $\mu_{TRGB}$ for NGC~4258), with a scatter of $\pm$0.052 mag.}
\label{compare}
\end{figure*}

Our independent TRGB measurements of stars in Cepheid hosts and NGC~4258 can be compared with those measured by the CCHP team. A comparison of the values of $m_{TRGB,0}$ with uncertainties\footnote{For the EDD values, we adopt the error on the determination of $m_{TRGB}$. The CCHP team does not explicitly provide the errors on their latest edge detection results, so we include their typical edge detection uncertainty of 0.03 mag on each point \citep{2019ApJ...882...34F}.} can be seen in the left-hand panel of Figure \ref{compare}. The mean difference between our two sets of measured values is $\Delta m_{TRGB,0}$ (EDD-CCHP) = 0.042 $\pm$ 0.012 mag with a scatter of $\pm0.038$ mag. 

However, we note that each team measures the tip in a different color range, so this comparison is not on equal terms. The CCHP tip is that of the bluest available RGB stars while the EDD methodology includes a broader and redder color range than those used by the CCHP. Since the $F814W$ TRGB is observed to \textit{decrease} in brightness at redder colors (with increasing metallicity), we would anticipate that the EDD TRGB measurements of higher metallicity populations would tend to be systematically fainter, which is in line with what is found. Ultimately, the translation of apparent TRGB magnitudes to distance moduli depend on the separate CCHP and EDD absolute magnitude calibrations and accounting for foreground obscuration. 

In anticipation of assessing the Hubble constant, let us factor out zero-point calibrations by comparing CCHP and EDD distances normalized to their respective distance estimates to the maser host NGC~4258. To derive values from the CCHP we begin by taking the most recent CCHP determinations of $\mu_{TRGB}$ \citep{2019ApJ...882...34F,2021arXiv210112232H} with an adjustment to the new zero-point of $M^{TRGB}_{F814W}$ = $-$4.045 \citep{2021arXiv210613337H}. We then subtract the CCHP distance modulus of NGC~4258 ($\mu$ = 29.392) from each of these values to obtain distance moduli that are relative to NGC~4258. Because the same TRGB absolute magnitude is assumed for the SN hosts and NGC 4258, its value is nullified in the difference between distance moduli.   

For the same galaxies, we obtain the TRGB distance moduli as reported in EDD, which include the zero-point and color-calibration of \cite{2007ApJ...661..815R}. From each, we subtract our new TRGB distance modulus for NGC~4258 ($\mu$ = 29.446) to obtain a set of distance moduli that are also relative to NGC~4258. The comparison between the CCHP and EDD values are shown in the right-hand panel of Figure \ref{compare}. The mean offset between the relative distance scales is $\Delta \mu$ (EDD-CCHP) = $-$0.028 $\pm$ 0.035 mag, or $\sim$1.5$\%$ in distance, with a \textbf{scatter} of $\pm0.052$ mag. The error on $\Delta \mu$ ($\pm$0.035 mag) is the quadrature summation of errors in each of the EDD and CCHP NGC~4258 tip values and the uncertainty in the differential between EDD and CCHP values of the tip absolute magnitude. It is important to note that this $-$0.028 mag difference is weighted by galaxy, and not by SN as is necessary for the determination of the Hubble constant.

An unmentioned detail is that the observed TRGB magnitudes must be corrected for foreground extinction. We, in common with the CCHP team, assume foreground extinction values given by \cite{2011ApJ...737..103S}. The one exception is that of NGC~5643. The galaxy NGC~5643 is located at the relatively low galactic latitude of $b = 15\degree$ where the dust maps, which have relatively coarse resolution, give uncertain attenuation estimates. To determine E(B-V) for NGC~5643, we use the displacement of the zero-age main sequence (ZAMS) to measure the absolute foreground reddening to the galaxy (see the example shown for ALFAZOAJ1952+1428 in \citealt{2017ApJ...835...78R}). Using this method, we find a value of E(B-V) = 0.161, slightly higher than the value of E(B-V) = 0.149 from \cite{2011ApJ...737..103S}. For the purposes of producing Figure \ref{compare} and the values and discussion presented in this subsection, we have adopted the \cite{2011ApJ...737..103S} value to ensure an even comparison. For the purposes of determining $H_{0}$ (in the next section), we adopt our new value of E(B-V) = 0.161 $\pm$ 0.024 mag.

To recap, the left-hand side of Figure \ref{compare} compares the CCHP and EDD apparent, extinction-corrected observed magnitudes of the TRGB ($m_{TRGB,0}$) for all the galaxies under discussion (including NGC~4258). We find that the EDD values are $\sim$2$\%$ fainter. Because each group measures the tip over a different color range (with EDD probing redder), we expect a systematic offset in the observed direction that originates solely from the different color ranges used to measure the TRGB. The right-hand side of Figure \ref{compare} shows the comparison between our two distance measures relative to NGC~4258 ($-$0.028 mag), and is the more relevant comparison for the purpose of setting the extragalactic distance scale, and hence the value of $H_{0}$. Recently, \cite{2021arXiv210613337H} presented a similar figure to the right-hand side of our Figure \ref{compare} in Figure 13 of that paper. An offset of 0.001 mag was found, smaller than the value we present here. Two differences are to be noted. The first is that we include our distance to NGC~1404, which was determined recently. The second is our new TRGB measurement for NGC~4258, which forms the basis of the zero-point scaling for the entire diagram. Broadly speaking, there is concordance that our two distance measures are in excellent agreement and that the TRGB is a reliable and efficient method of determining distances to nearby galaxies. 

\begin{deluxetable*}{lccccccccc}
\tablecaption{Values Used to Determine $H_{0}$}

\tablehead{\colhead{Host} & \colhead{$m_{TRGB}$}  & \colhead{$Color_{0}$} & \colhead{$A_{F814W}$} & \colhead{$\mu_{TRGB,cal}$} & \colhead{SN} & \colhead{$m_B^{Pan}$} & \colhead{$m_{B'}^{CSP}$}}

\startdata
M66           & 26.32  $\pm$ 0.03   & 1.58 $\pm$ 0.04   & 0.050   & 30.22  $\pm$ 0.05  & 1989B       & 10.98 $\pm$ 0.12              & 11.16 $\pm$ 0.07              \\ \hline
M96           & 26.30  $\pm$ 0.02   & 1.62 $\pm$ 0.06   & 0.038   & 30.20  $\pm$ 0.05  & 1998bu      & 11.00 $\pm$ 0.12              & 11.01 $\pm$ 0.06              \\ \hline
M101          & 25.08  $\pm$ 0.03   & 1.27 $\pm$ 0.03   & 0.013   & 29.08  $\pm$ 0.05  & 2011fe      & 9.83  $\pm$ 0.10              & 9.82  $\pm$ 0.03              \\ \hline
N1316         & 27.42  $\pm$ 0.05   & 1.47 $\pm$ 0.07   & 0.033   & 31.36  $\pm$ 0.07  & 1980N       & 11.98 $\pm$ 0.10              & 12.08 $\pm$ 0.06              \\ \hline
N1316         & 27.42  $\pm$ 0.05   & 1.47 $\pm$ 0.07   & 0.033   & 31.36  $\pm$ 0.07  & 1981D       & 11.61 $\pm$ 0.23              & 11.99 $\pm$ 0.17              \\ \hline
N1316         & 27.42  $\pm$ 0.05   & 1.47 $\pm$ 0.07   & 0.033   & 31.36  $\pm$ 0.07  & 2006dd      & 11.92 $\pm$ 0.08              & 12.38 $\pm$ 0.03              \\ \hline
N1365         & 27.43  $\pm$ 0.03   & 1.28 $\pm$ 0.04   & 0.031   & 31.41  $\pm$ 0.05  & 2012fr      & 11.93 $\pm$ 0.13              & 12.09 $\pm$ 0.03              \\ \hline
N1404         & 27.31  $\pm$ 0.06   & 1.35 $\pm$ 0.03   & 0.017   & 31.29  $\pm$ 0.07  & 2007on      & 12.46 $\pm$ 0.19              & 12.39 $\pm$ 0.07              \\ \hline
N1404         & 27.31  $\pm$ 0.06   & 1.35 $\pm$ 0.03   & 0.017   & 31.29  $\pm$ 0.07  & 2011iv      & 11.98 $\pm$ 0.15              & 12.03 $\pm$ 0.06              \\ \hline
N1448         & 27.35  $\pm$ 0.04   & 1.29 $\pm$ 0.03   & 0.021   & 31.33  $\pm$ 0.06  & 2001el      & 12.18 $\pm$ 0.09              & 12.30 $\pm$ 0.04              \\ \hline
N4038         & 27.72  $\pm$ 0.13   & 1.14 $\pm$ 0.07   & 0.071   & 31.68  $\pm$ 0.14  & 2007sr      & 12.30 $\pm$ 0.10              & 12.30 $\pm$ 0.15              \\ \hline
N4258         & 25.43  $\pm$ 0.03   & 1.32 $\pm$ 0.04   & 0.025   & --                 & --          & --                            & --                            \\ \hline
N4424         & 27.05  $\pm$ 0.05   & 1.38 $\pm$ 0.02   & 0.031   & 31.00  $\pm$ 0.07  & 2012cg      & 11.62 $\pm$ 0.12              & 11.72 $\pm$ 0.06              \\ \hline
N4526         & 27.03  $\pm$ 0.02   & 1.36 $\pm$ 0.03   & 0.035   & 30.99  $\pm$ 0.05  & 1994D       & 11.51 $\pm$ 0.17              & 11.76 $\pm$ 0.04              \\ \hline
N4536         & 27.03  $\pm$ 0.12   & 1.27 $\pm$ 0.04   & 0.028   & 31.01  $\pm$ 0.13  & 1981B       & 11.60 $\pm$ 0.09              & 11.64 $\pm$ 0.04              \\ \hline
N5643         & 26.70  $\pm$ 0.03   & 1.30 $\pm$ 0.03   & 0.278   & 30.42  $\pm$ 0.07  & 2013aa      & 11.27 $\pm$ 0.14              & 11.31 $\pm$ 0.09              \\ \hline
N5643         & 26.70  $\pm$ 0.03   & 1.30 $\pm$ 0.03   & 0.278   & 30.42  $\pm$ 0.07  & 2017cbv     & 11.20 $\pm$ 0.14              & 11.30 $\pm$ 0.09              \\ \hline
\enddata

\tablecomments{Table providing the data used to measure the value of $H_{0}$ using our TRGB determinations and both the Pantheon and Carnegie Supernova Project (CSP) sample of supernovae. The individual columns are 1) host galaxy, 2) the measurement of $m_{TRGB}$ in F814W, 3) the reddening corrected median F606W-F814W color of the measured TRGB, 4) adopted extinction for F814W, 5) the distance modulus to each galaxy, after re-scaling the \cite{2007ApJ...661..815R} zero-point with our new TRGB measurement in NGC~4258,  6) name of the individual type Ia supernova, 7) standardized peak magnitude of the supernova provided by the Pantheon supernova sample, and 8) standardized peak magnitude of the supernova provided by the CSP sample. Note that the error as given in 8) does not include an intrinsic error, and so we add 0.10 mag in quadrature during the process of determining $H_{0}$.}
\label{tab:H0}
\end{deluxetable*}

\section{The Hubble Constant}

With our independently determined TRGB distances to the CCHP sample of galaxies, along with a new and consistent calibration of the TRGB enabled by the new observations in NGC~4258, we can derive the value of the Hubble Constant. Our procedure is relatively straightforward, as we are \textit{not} reassessing the SN Ia portion of the distance ladder. Rather, we are reusing the two mostly widely used standardized SN Ia samples. These are the Pantheon \citep{2018ApJ...859..101S} and Carnegie Supernova Project (CSP, \citealt{2006PASP..118....2H,2017AJ....154..211K}) samples of supernovae.  Each of the two is assessed separately, and the relevant parameters are provided in Table \ref{tab:H0}.

We make use of the following Equation from \cite{2016ApJ...826...56R}:
\begin{equation}
    {\rm log_{10}}(H_{0}) = \frac{M^{0}_{B} + 5a_{b} +25}{5}
\end{equation}
where $a_b$ is the intercept of the SN Ia Hubble diagram (translated to $z=0$) and $M^{0}_{B}$ is the standardized absolute luminosity of a type Ia supernovae (at the fiducial point of a specific methodology for standardizing SN Ia light curves, e.g., \cite{2018ApJ...859..101S} or \cite{2018ApJ...869...56B}. $M^{0}_{B}$ may be calibrated from our TRGB distances in SN Ia hosts. 

We obtain an initial value of $\mu_{TRGB}$ for each galaxy by following Equation 2, the zero-point and color calibration from \cite{2007ApJ...661..815R}. We then rescale these distances based on our new measurement of the TRGB in NGC 4258. In other words, we take the values of $\mu_{TRGB}$ for each galaxy and adjust these distance moduli by the difference in our measured value of $\mu_{TRGB}$ for NGC~4258 based on the \cite{2007ApJ...661..815R} calibration ($\mu$ = 29.446) and the known value based on the megamaser distance ($\mu$ = 29.397, \citealt{2019ApJ...886L..27R}). We refer to these newly calibrated set of distance moduli as $\mu_{TRGB}^{cal}$, where for each galaxy:

\begin{equation}
     \mu_{TRGB}^{cal} = \mu_{TRGB} - \mu_{TRGB}^{N4258} + \mu_{Maser}^{N4258}. 
\end{equation}

To obtain $M^{0}_{B}$, we take the weighted average of the individual values of $M^{0}_{B}$ for each supernova. These are found by $M^{0}_{B}$ = $m_{B}-\mu_{TRGB,cal}$, where $m_{B}$ is the standardized peak magnitude of each type Ia supernovae in the sample. The weights used in this procedure are the errors in the measured TRGB distances added in quadrature with the errors in $m_{B}$. With the value of $M^{0}_{B}$ in hand, we adopt the value of $a_{b}$ measured from supernovae in the Hubble flow and corrected to $z=0$ (with $q_0=-0.55$ and $j_0=1$ as in \citealt{2016ApJ...826...56R} and \citealt{2019ApJ...882...34F}). For the Pantheon sample (with SN peak magnitudes $m_B^{Pan}$), we use $a_{b}$ = \abPan \ which was derived from 746 SNe Ia at $0.023 < z < 0.15$ in the expanded Pantheon sample that pass standard quality cuts\footnote{A few of the SN Ia in the TRGB hosts do not pass the Pantheon quality cuts, i.e., SN 1989B and 1998bu have $A_V \sim 1.0$, while SN 1994D, 2007on and 2011iv are transitional fast-decliners and SN 1981D is poorly observed. They are all retained to be consistent with the F19 CCHP analysis.} \citep{2018ApJ...859..101S, 2022arXiv220204077B}. While the CCHP program did not explicitly provide the equivalent $a_b$ CSP value in \cite{2019ApJ...882...34F}, we are able to derive $a_{b}$ = \abCSP \ from their values of $H_0=69.8$~km/s/Mpc and $M^{0}_{B}=-19.23$ using Equation 3.

Along with the measurement of $H_{0}$, equally important is the error budget associated with our final reported values. To provide the final quoted uncertainty associated with our result, we add in quadrature the following terms: 1) the mean error in $M^{0}_{B}$ (which accounts for the errors from the individual TRGB (including an assumed 15$\%$ uncertainty in the extinction) and supernova measurements), 2) the error in the NGC~4258 TRGB measurement, 3) the error in the NGC~4258 megamaser distance, and 4) the error in 5$a_{b}$. These individual contributions are also summarized in Table \ref{tab:H0errors}.

\begin{deluxetable}{lcc}

\tablecaption{Sources of Error in $H_{0}$ (in magnitudes)}
\tablehead{\colhead{Error Term} & \colhead{Value (Pan.)} & \colhead{Value (CSP)}} 

\startdata
SN--TRGB Linkage & 0.037 & 0.034 \\
$\,$ (Mean 16 SNe)  &  0.035 & 0.032 \\
$\,$  (Mean 11 TRGB) & 0.010 & 0.010 \\
NGC~4258 -- $m_{TRGB}$  & 0.027 & 0.027 \\
NGC~4258 -- $\mu_{maser}$  & 0.032 & 0.032 \\
SN Ia--Hubble Intercept & 0.005 & 0.010 \\
\hline
\textbf{Quadrature Sum} & \textbf{0.056} & \textbf{0.055} 
\enddata

\tablecomments{A breakdown of the sources of error in our determination of $H_{0}$, using alternatively the Pantheon and CSP samples of supernovae. Descriptions of the individual terms are 1) the error in the SN-TRGB linkage, or $M^{0}_{B}$ (broken down into its two constituent components), 2) the error in the value of $m_{TRGB}$ for NGC~4258, 3) the error in the megamaser distance modulus ($\mu$) for NGC~4258 \citep{2019ApJ...886L..27R}, and 4) the error in the SN Ia Hubble diagram intercept, or 5$a_{b}$.}
\label{tab:H0errors}

\end{deluxetable}

Following the above prescription, from the Pantheon sample we find a value of $M^{0}_{B}$ = \MBPan\ and $H_{0}$~= \hubblePan. Using the CSP sample, we find $M^{0}_{B}$ = \MBCSP\ and the same value of $H_{0}$ = \hubbleCSP, implying good agreement between the values of $H_{0}$ derived from the two supernova samples.  We note it is not meaningful to compare the values of $M^{0}_{B}$ and $a_b$ between the two SN Ia light curve standardization methods because each depends on fiducial terms (i.e., which SN Ia light curve shape and point along its light curve serves as the reference). This difference cancels in Equation 3 and the determination of $H_0$.  We also calculated $H_0$ accounting for the covariance of multiple SNe Ia in the same host using the same TRGB distance (i.e., 3 in NGC 1316, 2 in NGC 1404 and 2 in NGC 5643) and found it made little difference-- the net effect was raising the Pantheon value by $\sim$ 0.1 km/s/Mpc and lowering the CCHP value by the same. We retain the simpler calculation for its accessibility.

Our preferred value of $H_{0}$ (\hubblePan) is somewhat larger than the value of $H_{0}$ = 69.8 $\pm$ 0.6 (stat) $\pm$ 1.6 (sys) km/s/Mpc found by \cite{2021arXiv210615656F} due to differences that are summarized in Table \ref{tab:magDifferences}. We provide these differences in magnitudes which are more directly traceable to the observations and correspond to units of $5\log (H_0)$, where a difference of 0.03 mag corresponds to a difference of $\sim$1 km/sec/Mpc. Broken down explicitly, these differences are:

\begin{enumerate}
    \item There is a $-$0.028 mag difference (EDD$-$CCHP) between our two distance scales, when compared relative to the maser host NGC~4258 (weighted by galaxy). There are multiple factors which contribute to this (albeit small) difference. Our measurement of the TRGB in NGC~4258 is slightly further (0.055 mag) than the average difference between our two sets of measurements of $m_{TRGB}$ for the sample as a whole (0.042 mag). When weighted by the SNe, this difference is $-$0.02 mag for \cite{2021arXiv210615656F} or $-$0.03 mag for \cite{2019ApJ...882...34F}. Another difference is the assumed color structure of the TRGB.  The CCHP assume a slope of zero for their chosen color ranges, while we assume the slope found by \cite{2007ApJ...661..815R}. Because the TRGB sample brackets the color of NGC 4258, the net effect of the color term is an additional 0.01 mag in $H_0$. A shallower or steeper color term would produce a fraction of this 0.01 mag difference.
    
    \item We remove four galaxies (NGC~1309, NGC~3021, NGC~3370, and NGC~5584) for which we are not able to determine reliable TRGB distances (see $\S$4.2 for details). Comparing the error-weighted mean of the calibrated SN luminosities in \cite{2019ApJ...882...34F} and \cite{2021arXiv210615656F} with and without these four, there is no change to $M_B$ (or $H_0$) however, the error increases by 10\%.
    
    \item Line 4 in Table \ref{tab:magDifferences} provides differences in $H_0$ related to the host NGC 5643.  NGC 5643 hosted two SNe Ia, 2017cbv and 2013aa which were not included in \cite{2019ApJ...882...34F} (raising $H_0$ by 0.01 mag here), but were included in \cite{2021arXiv210615656F} (no additional difference). If we use the foreground extinction estimate, $A_I$, from \cite{2011ApJ...737..103S} for NGC 5643 which is 0.02 mag lower than our own as done by F21 our value of $H_0$ would decrease by 0.07 km s${-1}$ Mpc$^{-1}$.
    
    Line 3 in Table \ref{tab:magDifferences} provides differences related to the host NGC 1404. NGC 1404 hosted SN 2007on and 2011iv, both used in the CCHP result from \cite{2019ApJ...882...34F} and \cite{2020ApJ...891...57F}. These SNe Ia are fast decliners or ``transitional'' SNe ($S_{BV}=-0.6$) and after standardization they differ from each other by 0.4 mag. SN 2007on is $\sim$0.2 mag fainter than the mean $M_B$ of 18 SNe~Ia and conversely SN 2011iv is $\sim$0.2 mag brighter \citep{2019ApJ...882...34F}, yet both are consistent with the mean at $\sim2\sigma$. \cite{2021arXiv210615656F} now excludes SN 2007on on the grounds that ``it appears significantly underluminous''.  SN 2007on does not appear to be an outlier in our analyses using Chauvenet's criterion. It is fainter than the mean supernova $M_B$ by 2.0$\sigma$ for the CSP SN analysis and by $1.9 \sigma$ for the Pantheon SN analysis, with one other SN more discrepant and in the other direction (SN 1981D)\footnote{Hoyt et al. (2021) suggested SN 2007on could be subluminous by comparing it to just the SNe in NGC 1316 and assuming it to be at a similar distance as NGC 1404. However, as stated, SN 1981D is brighter than the mean in the Pantheon analysis by greater significance than 2007on is fainter than the mean. All 3 supernovae in NGC 1316 are brighter than the full sample mean, making the difference appear greater than when comparing to the whole sample.}. For consistency, we would need to exclude both SNe in this host, all transitional SNe in the calibrator and Hubble flow set, and all SNe from either analysis that are this discrepant. The alternative is to exclude none, which is what we have done. Retaining 2007on raises $H_0$ by 0.01 mag relative to \cite{2021arXiv210615656F}. If we remove 2007on from our sample as done by F21 our value of $H_0$ would decrease by $\sim$0.5 km s${-1}$ Mpc$^{-1}$.

    \end{enumerate}
    
\begin{deluxetable}{lcc}

\tablecaption{Sources of Differences in $H_{0}$ Between EDD and CCHP (in magnitudes)}
\tablehead{\colhead{Term} & \colhead{$\Delta$ F19} & \colhead{$\Delta$ F21}} 

\startdata
Zero-point (NGC~4258) & 0.06 & 0.06 \\
No TRGB Fits (Four Hosts) & 0.00 & 0.00 \\
NGC 1404 & 0.01 & 0.01\\
NGC 5643* & 0.01 & -- \\
$\Delta m_{TRGB}$* & -0.03. & -0.02 \\
\hline
\textbf{Total} & \textbf{0.05} & \textbf{0.05} \\
\enddata

\tablecomments{A breakdown in the sources of differences in $H_0$ (in magnitudes) between the EDD and CCHP results. ``*" indicates these values are weighted by supernovae, and not by host. $\Delta$F19 and $\Delta$F21 refer to the differences between \cite{2019ApJ...882...34F} and EDD, and \cite{2021arXiv210615656F} and EDD, respectively. Descriptions of the individual terms are 1) the measured zero-point calibration of the TRGB in the megamaser host galaxy NGC~4258, 2) the lack of measurements for four host galaxies in EDD (see $\S$4.2 for details), 3) the inclusion of NGC~1404 (not directly measured in F19, and we include SN 2007on which F19 included and F21 excluded), 4) the inclusion of NGC~5643 (not available for F19), and 5) the mean difference in measured values for the TRGB of the remaining hosts.  }
\label{tab:magDifferences}

\end{deluxetable}

To directly compare the effects of sample selection versus TRGB measurements themselves, we can choose to hew as close as possible to the choices made by the CCHP. Namely, if we remove SN 2007on from our sample (however see point 3 above) and revert our extinction measurement for NGC~5643 to those obtained from the dust maps, we find $H_{0}$ = 70.84 km/s/Mpc using the CSP sample of supernovae, as opposed to our primary measurement of $H_0$ = \hubblePan. We can also compare our final values with the result derived from Cepheids and NGC 4258 in \cite{2019ApJ...886L..27R}, namely $H_{0}$ = 72.0 $\pm$ 1.9 km/s/Mpc. The relative difference is 0.5 $\pm$ 2.0 km/s/Mpc (excluding the maser distance uncertainty in common to both). We can separate the terms in the error budget in Table \ref{tab:H0errors} which are independent from this Cepheid-SN Ia measurement. These include the the tip measurement for NGC 4258 and 10 of the 16 SNe Ia which were not included in the SH0ES analyses. The TRGB-SN Ia error {\it independent} of the Cepheid-SN Ia ladder using NGC 4258 is 0.035 mag or 1.1 km/s/Mpc, thus shows good agreement between the TRGB and Cepheid routes. We also note that a detailed analysis of the covariance between calibrator and Hubble flow SNe Ia by \cite{2020ApJ...894...54D} not included here shows such covariance may raise $H_0$ by 0.5 km/s/Mpc in the Pantheon SN analysis (see their Table 2). This effect has not yet been measured for SNe Ia in TRGB hosts nor is this information available for the CSP SN analysis.

\begin{figure}
\epsscale{1.15}
\plotone{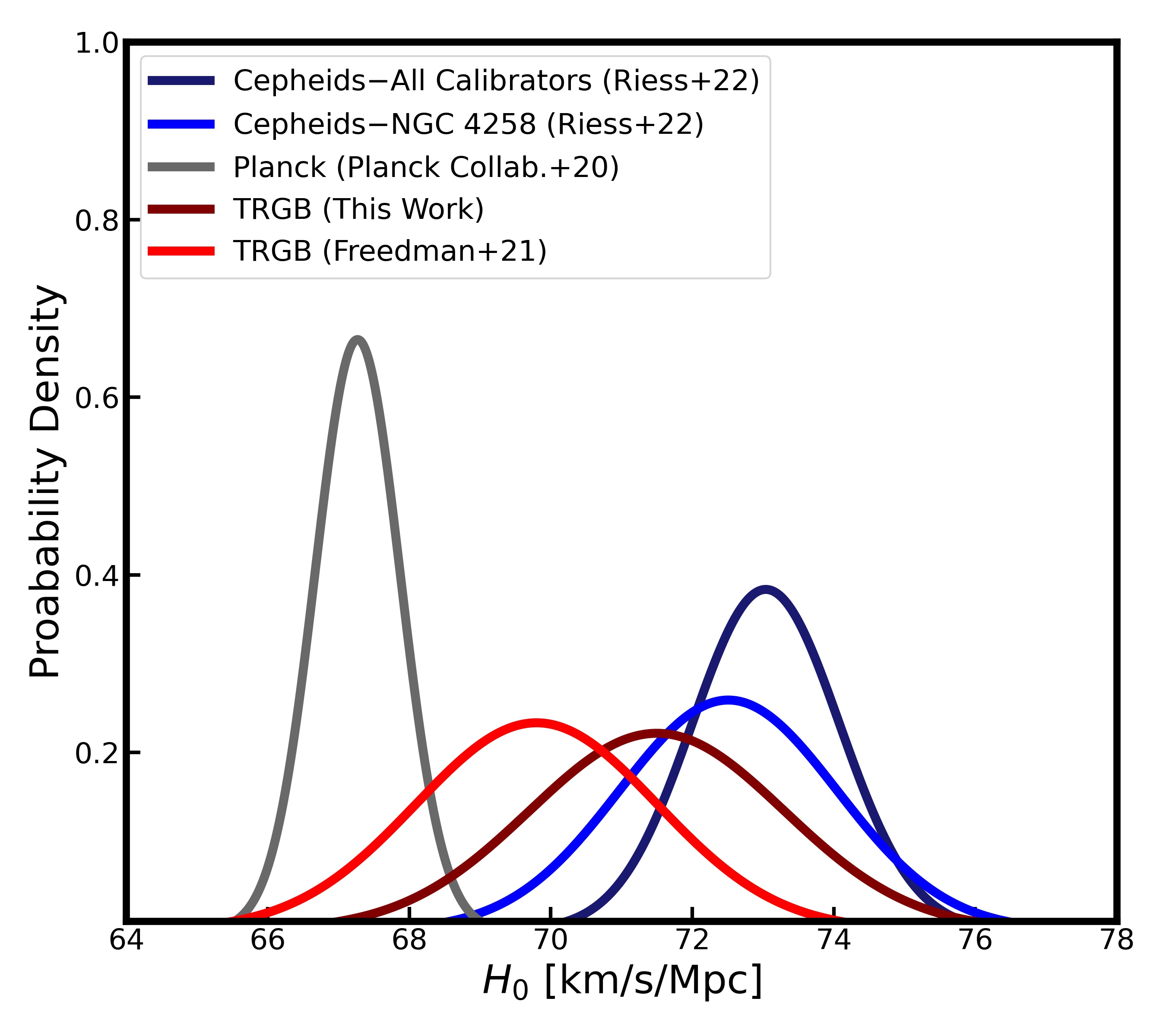}
\caption{Probability density functions (assuming Gaussian distributions) for the Hubble Constant as determined by Planck \citep{2020A&A...641A...6P}, Cepheid variables \citep{2021arXiv211204510R}, the CCHP via TRGB \citep{2021arXiv210615656F}, and our results from the TRGB. For the results from Cepheids, we plot both the main result from SH0ES, as well as one which uses only NGC~4258 as a zero-point calibrator. Ultimately we see that results from the TRGB fall in between of those from Planck and the Cepheids.}
\label{summaryH0}
\end{figure}

\section{Summary and Future Prospects}

In this paper, we present a new determination of the Hubble Constant using TRGB distances as calibrators for type Ia supernovae. We use the same underlying data as \cite{2019ApJ...882...34F} and \cite{2021arXiv210615656F}, except we remove the four most distant hosts (NGC~1309, NGC~3021, NGC~3370, and NGC~5584) as we find the data to be insufficient to derive a robust measurement of the TRGB. We also include two hosts (NGC~1404 and NGC~5643, with a total of 4 SNe Ia) from \cite{2021arXiv210112232H} and \cite{2021arXiv210615656F}. Combining these results with a new measurement of the TRGB in the maser host NGC~4258 based on new HST observations and the Pantheon or CSP sample of supernovae, we find a value of the Hubble Constant of $H_{0}$ = \hubblePan\ (for both supernova samples). This value is $\sim$2$\%$ larger than the CCHP value of $H_{0}$ = 69.8 $\pm$ 0.8 $\pm$ 1.7 km/s/Mpc found by \cite{2019ApJ...882...34F}. Our determination of $H_0$ from the EDD TRGB lies near the mid point between the value from the CCHP TRGB and the most recent SH0ES Cepheid-based result of 73.0 km/s/Mpc \citep{2021arXiv211204510R}. The situation can be seen summarized in Figure \ref{summaryH0}, where ultimately, both our and the CCHP's determination of the Hubble Constant from the TRGB-SNIa ladder fall in between those of Planck and the Cepheids. Key sources of uncertainty in the TRGB distance ladder remain, including the zero-point and color-calibrations for the absolute magnitude of the TRGB, as well as the restricted number of TRGB calibrators for type Ia supernovae.

In the coming months, we anticipate another important measurement of the Hubble Constant, this time from Cosmicflows-4 (Tully et al., in prep). Cosmicflows is a program that collects tens of thousands of redshift-independent distances to galaxies, placed on a common distance scale. Cosmicflows-4 will provide $\sim$50,000 galaxy distances, and includes TRGB distances from the CMDs/TRGB catalog as a key calibrator of methods that extend to greater distance (e.g. the Tully-Fisher relation, type Ia supernovae, etc.). 

In the coming years, the sample of galaxies with TRGB distances and type Ia supernova will continue to expand. Part of this increase is due to additional programs with \textit{HST}, but the truly substantial increase will happen with \textit{JWST}. With \textit{HST}, we have shown that reliable TRGB distances can be measured out to $\sim$20~Mpc. With \textit{JWST}, this limit will at least double, providing a substantial increase in volume and thus the number of calibrators. Additionally, \textit{JWST} will allow for the TRGB calibration of a substantial number of galaxies whose distances can also be measured via surface-brightness fluctuations \citep{2021ApJ...911...65B}, providing a truly independent measure of $H_{0}$ from the current Cepheid+Type Ia supernova distance ladder.


\begin{acknowledgements}

This research is supported by an award from the Space Telescope Science Institute in support of program SNAP-15922. This research has made use of the NASA/IPAC Extragalactic Database (NED), which is operated by the Jet Propulsion Laboratory, California Institute of Technology, under contract with the National Aeronautics and Space Administration.  G.A. thanks Zach Claytor and Ryan Dungee for useful discussions. A.R. thanks Dan Scolnic and Dillon Brout for useful discussions about SN light curve fitting. We thank the anonymous referee for useful comments which improved the quality of our manuscript.

\end{acknowledgements}

\facilities{\textit{HST}: ACS/WFC $\&$ WFC3/UVIS}

\software{DOLPHOT \citep{2000PASP..112.1383D, 2016ascl.soft08013D}, \textit{DrizzlePac} \citep{2015ASPC..495..281A}}

\appendix
\section{Figure Set for CCHP Targeted Galaxies}

\begin{figure}
\epsscale{1.10}
\plotone{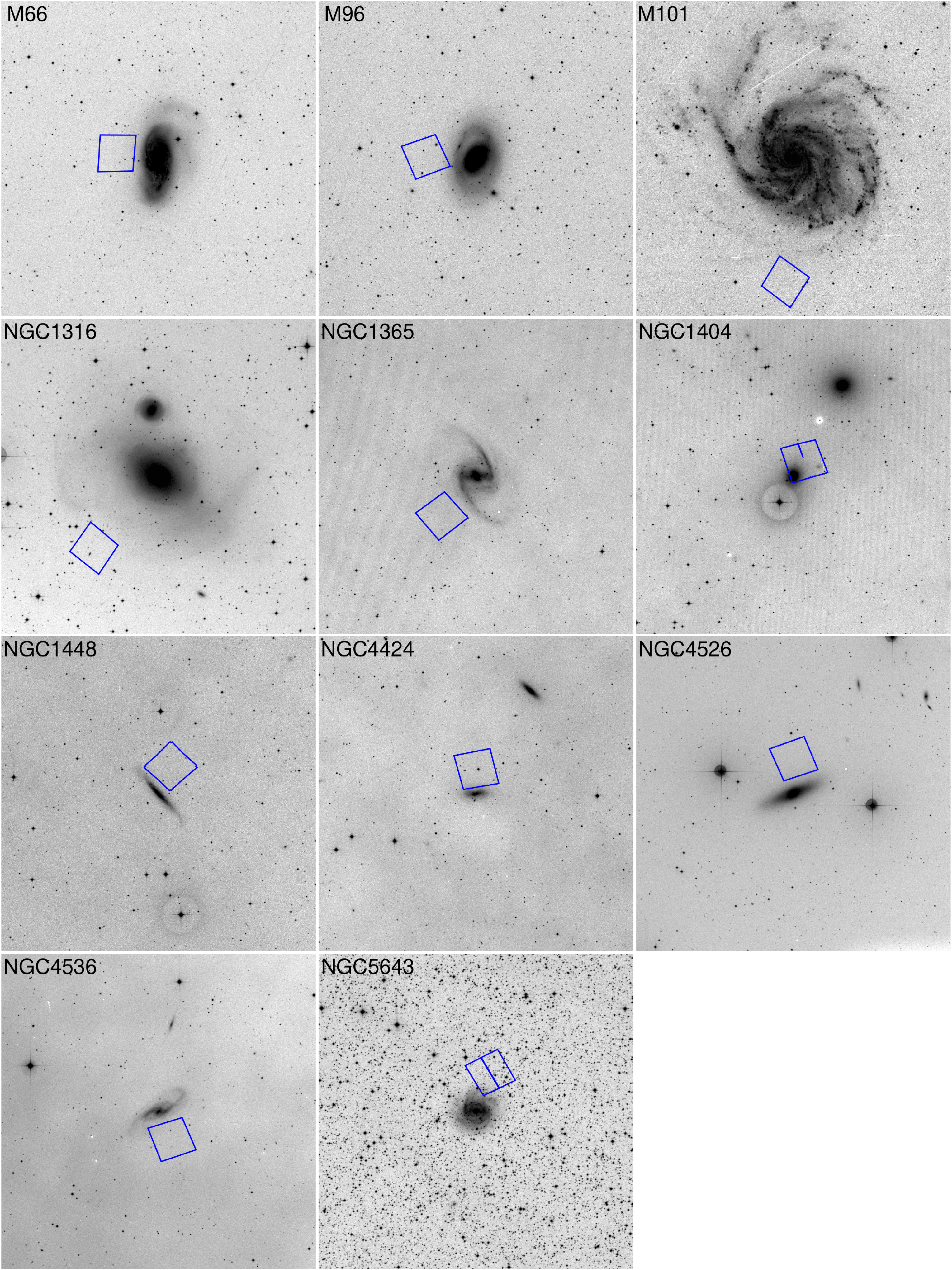}
\caption{Digitzed Sky Survey images for each of the CCHP targeted galaxies, with the HST/ACS footprints of the observations from GO-13691 and GO-15642 overlaid in blue. Each DSS image is 30$\times$30 arcminutes in size.}
\label{footprintCollage}
\end{figure}

\begin{figure}
\epsscale{1.15}
\plotone{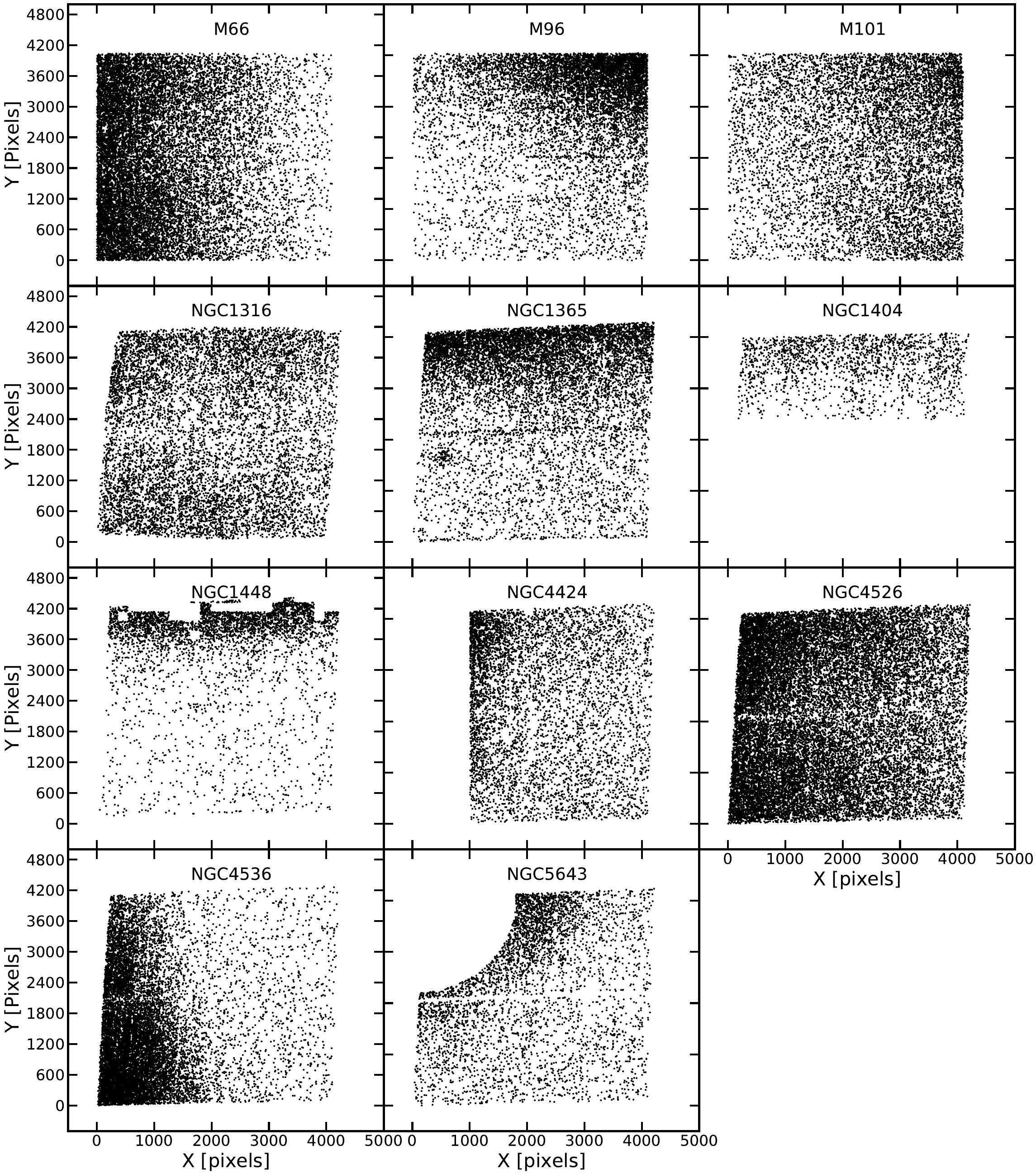}
\caption{Plots of stars which pass our DOLPHOT quality cuts for each of the CCHP targeted galaxies. For four galaxies (NGC~1404, NGC~1448, NGC~4424, and NGC~5643), we trimmed away a portion of the stellar field to exclude regions with extreme levels of crowding or visible star-formation.}
\label{starCollage}
\end{figure}

\begin{figure}
\epsscale{1.15}
\plotone{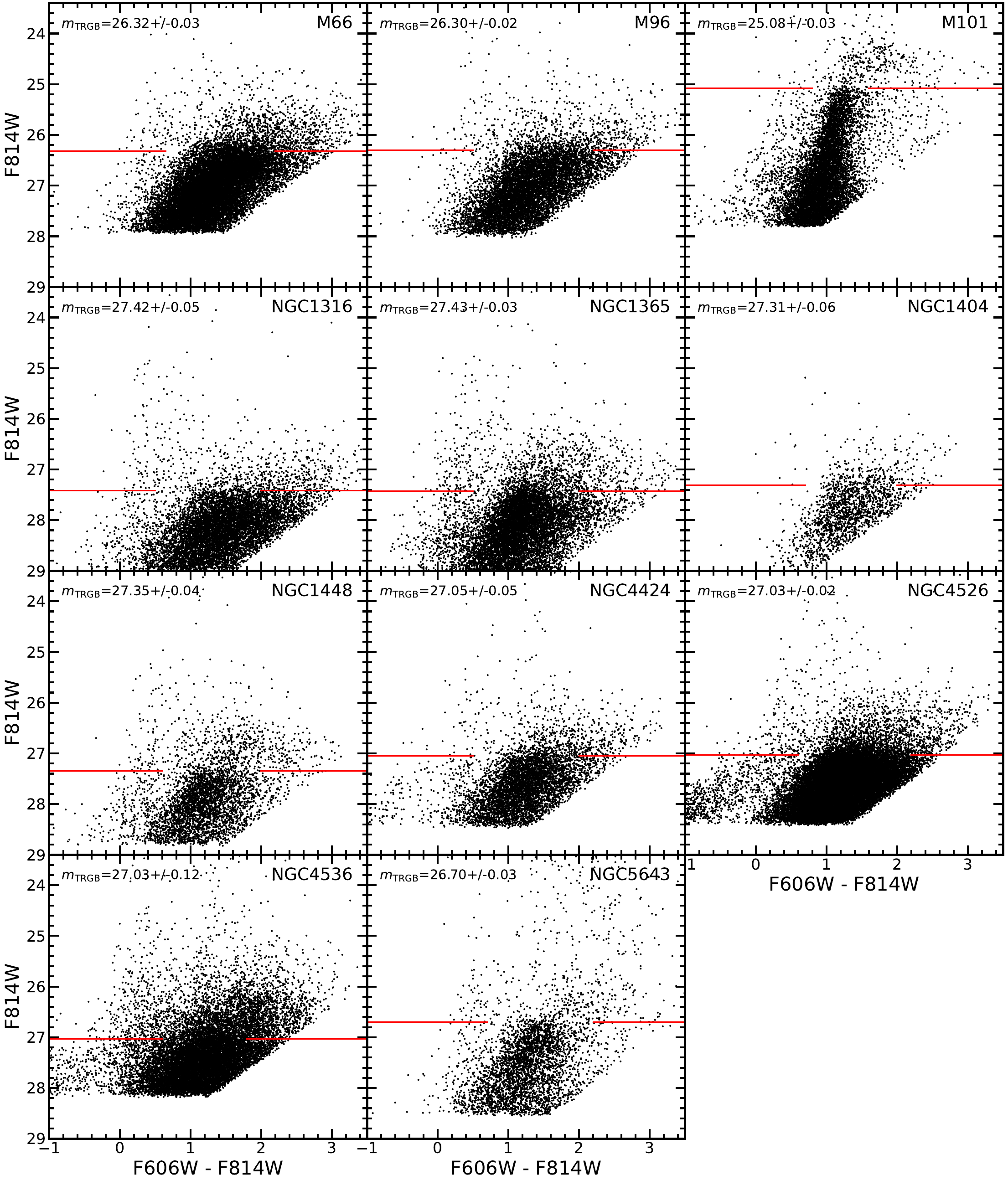}
\caption{Color magnitude diagrams showing our reduction of the data obtained by the CCHP (GO-13691 and GO-15642). For NGC~1404, NGC~1448, NGC~4425 and NGC~5643, these color-magnitude diagrams are spatially selected to exclude regions of extreme crowding or visible star-formation (see Figure \ref{starCollage}). Our determinations of the TRGB are shown with red lines and labelled in the top left of each diagram. The gap in the broken red line shows the color range used for our determination of the TRGB.}
\label{CMDcollage}
\end{figure}

\begin{figure}
\epsscale{1.15}
\plotone{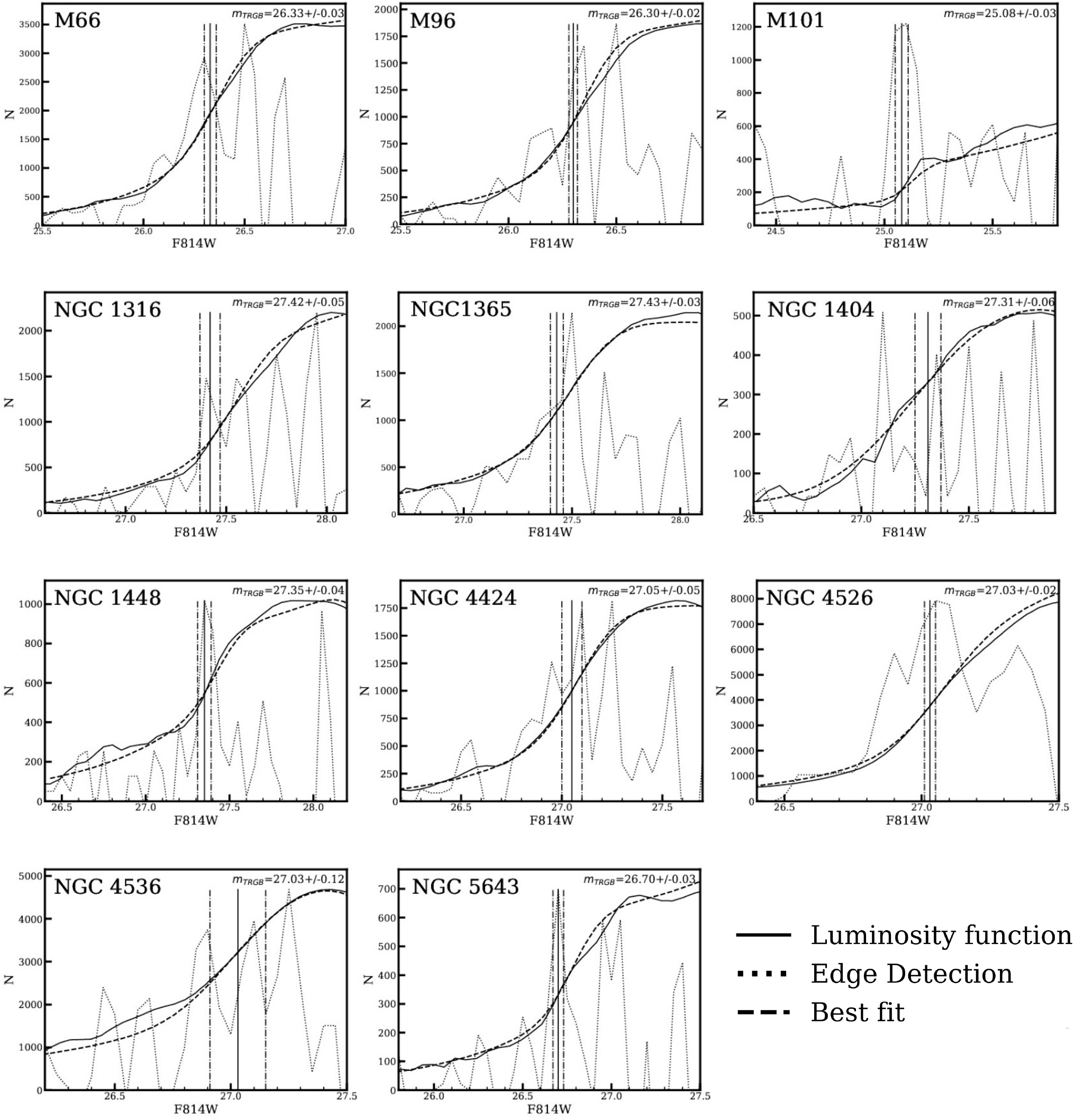}
\caption{Observed luminosity functions, best-fit luminosity functions, and edge-detection results (shown only for visual comparison) corresponding to the color-magnitude diagrams shown in Figure \ref{CMDcollage}. Note that while some of our best-fit results correspond to clean edge detections in our data (e.g. M101), other CMDs have edge-detection responses which are much noisier in our data (e.g. NGC~4424).}
\label{fig:LFcollage}
\end{figure}

\bibliography{paper}

\begin{thebibliography}{}
\expandafter\ifx\csname natexlab\endcsname\relax\def\natexlab#1{#1}\fi
\providecommand{\url}[1]{\href{#1}{#1}}

\bibitem[{{Anand} {et~al.}(2019){Anand}, {Tully}, {Rizzi}, {Shaya}, \&
  {Karachentsev}}]{2019ApJ...880...52A}
{Anand}, G.~S., {Tully}, R.~B., {Rizzi}, L., {Shaya}, E.~J., \& {Karachentsev},
  I.~D. 2019, \apj, 880, 52

\bibitem[{{Anand} {et~al.}(2021{\natexlab{a}}){Anand}, {Lee}, {Van Dyk},
  {Leroy}, {Rosolowsky}, {Schinnerer}, {Larson}, {Kourkchi}, {Kreckel},
  {Scheuermann}, {Rizzi}, {Thilker}, {Tully}, {Bigiel}, {Blanc}, {Boquien},
  {Chandar}, {Dale}, {Emsellem}, {Deger}, {Glover}, {Grasha}, {Groves},
  {Klessen}, {Kruijssen}, {Querejeta}, {S{\'a}nchez-Bl{\'a}zquez}, {Schruba},
  {Turner}, {Ubeda}, {Williams}, \& {Whitmore}}]{2021MNRAS.501.3621A}
{Anand}, G.~S., {Lee}, J.~C., {Van Dyk}, S.~D., {et~al.} 2021{\natexlab{a}},
  \mnras, 501, 3621

\bibitem[{{Anand} {et~al.}(2021{\natexlab{b}}){Anand}, {Rizzi}, {Tully},
  {Shaya}, {Karachentsev}, {Makarov}, {Makarova}, {Wu}, {Dolphin}, \&
  {Kourkchi}}]{2021arXiv210402649A}
{Anand}, G.~S., {Rizzi}, L., {Tully}, R.~B., {et~al.} 2021{\natexlab{b}}, arXiv
  e-prints, arXiv:2104.02649

\bibitem[{{Anderson}(2018)}]{2018wfc..rept...14A}
{Anderson}, J. 2018, {Focus-Diverse PSFs for Five Commonly Used WFC3/UVIS
  Filters}, Space Telescope WFC Instrument Science Report, ,

\bibitem[{{Anderson} \& {Ryon}(2018)}]{2018acs..rept....4A}
{Anderson}, J., \& {Ryon}, J.~E. 2018, {Improving the Pixel-Based
  CTE-correction Model for ACS/WFC}, Instrument Science Report ACS 2018-04, ,

\bibitem[{{Avila} {et~al.}(2015){Avila}, {Hack}, {Cara}, {Borncamp}, {Mack},
  {Smith}, \& {Ubeda}}]{2015ASPC..495..281A}
{Avila}, R.~J., {Hack}, W., {Cara}, M., {et~al.} 2015, in Astronomical Society
  of the Pacific Conference Series, Vol. 495, Astronomical Data Analysis
  Software an Systems XXIV (ADASS XXIV), ed. A.~R. {Taylor} \& E.~{Rosolowsky},
  281

\bibitem[{{Beaton} {et~al.}(2016){Beaton}, {Freedman}, {Madore}, {Bono},
  {Carlson}, {Clementini}, {Durbin}, {Garofalo}, {Hatt}, {Jang}, {Kollmeier},
  {Lee}, {Monson}, {Rich}, {Scowcroft}, {Seibert}, {Sturch}, \&
  {Yang}}]{2016ApJ...832..210B}
{Beaton}, R.~L., {Freedman}, W.~L., {Madore}, B.~F., {et~al.} 2016, \apj, 832,
  210

\bibitem[{{Beaton} {et~al.}(2018){Beaton}, {Bono}, {Braga}, {Dall'Ora},
  {Fiorentino}, {Jang}, {Mart{\'\i}nez-V{\'a}zquez}, {Matsunaga}, {Monelli},
  {Neeley}, \& {Salaris}}]{2018SSRv..214..113B}
{Beaton}, R.~L., {Bono}, G., {Braga}, V.~F., {et~al.} 2018, \ssr, 214, 113

\bibitem[{{Beaton} {et~al.}(2019){Beaton}, {Seibert}, {Hatt}, {Freedman},
  {Hoyt}, {Jang}, {Lee}, {Madore}, {Monson}, {Neeley}, {Rich}, \&
  {Scowcroft}}]{2019ApJ...885..141B}
{Beaton}, R.~L., {Seibert}, M., {Hatt}, D., {et~al.} 2019, \apj, 885, 141

\bibitem[{{Bennet} {et~al.}(2021){Bennet}, {Sand}, {Crnojevi{\'c}}, {Weisz},
  {Caldwell}, {Guhathakurta}, {Hargis}, {Karunakaran}, {Mutlu-Pakdil},
  {Olszewski}, {Salzer}, {Seth}, {Simon}, {Spekkens}, {Stark}, {Strader},
  {Tollerud}, {Toloba}, \& {Willman}}]{2021arXiv210108270B}
{Bennet}, P., {Sand}, D.~J., {Crnojevi{\'c}}, D., {et~al.} 2021, arXiv
  e-prints, arXiv:2101.08270

\bibitem[{{Blakeslee} {et~al.}(2021){Blakeslee}, {Jensen}, {Ma}, {Milne}, \&
  {Greene}}]{2021ApJ...911...65B}
{Blakeslee}, J.~P., {Jensen}, J.~B., {Ma}, C.-P., {Milne}, P.~A., \& {Greene},
  J.~E. 2021, \apj, 911, 65

\bibitem[{{Brout} {et~al.}(2022){Brout}, {Scolnic}, {Popovic}, {Riess},
  {Zuntz}, {Kessler}, {Carr}, {Davis}, {Hinton}, {Jones}, {Kenworthy},
  {Peterson}, {Said}, {Taylor}, {Ali}, {Armstrong}, {Charvu}, {Dwomoh},
  {Palmese}, {Qu}, {Rose}, {Stubbs}, {Vincenzi}, {Wood}, {Brown}, {Chen},
  {Chambers}, {Coulter}, {Dai}, {Dimitriadis}, {Filippenko}, {Foley}, {Jha},
  {Kelsey}, {Kirshner}, {M{\"o}ller}, {Muir}, {Nadathur}, {Pan}, {Rest},
  {Rojas-Bravo}, {Sako}, {Siebert}, {Smith}, {Stahl}, \&
  {Wiseman}}]{2022arXiv220204077B}
{Brout}, D., {Scolnic}, D., {Popovic}, B., {et~al.} 2022, arXiv e-prints,
  arXiv:2202.04077

\bibitem[{{Burns} {et~al.}(2018){Burns}, {Parent}, {Phillips}, {Stritzinger},
  {Krisciunas}, {Suntzeff}, {Hsiao}, {Contreras}, {Anais}, {Boldt}, {Busta},
  {Campillay}, {Castell{\'o}n}, {Folatelli}, {Freedman}, {Gonz{\'a}lez},
  {Hamuy}, {Heoflich}, {Krzeminski}, {Madore}, {Morrell}, {Persson}, {Roth},
  {Salgado}, {Ser{\'o}n}, \& {Torres}}]{2018ApJ...869...56B}
{Burns}, C.~R., {Parent}, E., {Phillips}, M.~M., {et~al.} 2018, \apj, 869, 56

\bibitem[{{Carretta} {et~al.}(2000){Carretta}, {Gratton}, {Clementini}, \&
  {Fusi Pecci}}]{2000ApJ...533..215C}
{Carretta}, E., {Gratton}, R.~G., {Clementini}, G., \& {Fusi Pecci}, F. 2000,
  \apj, 533, 215

\bibitem[{{Crnojevi{\'c}} {et~al.}(2019){Crnojevi{\'c}}, {Sand}, {Bennet},
  {Pasetto}, {Spekkens}, {Caldwell}, {Guhathakurta}, {McLeod}, {Seth}, {Simon},
  {Strader}, \& {Toloba}}]{2019ApJ...872...80C}
{Crnojevi{\'c}}, D., {Sand}, D.~J., {Bennet}, P., {et~al.} 2019, \apj, 872, 80

\bibitem[{{Dalcanton} {et~al.}(2009){Dalcanton}, {Williams}, {Seth}, {Dolphin},
  {Holtzman}, {Rosema}, {Skillman}, {Cole}, {Girardi}, {Gogarten},
  {Karachentsev}, {Olsen}, {Weisz}, {Christensen}, {Freeman}, {Gilbert},
  {Gallart}, {Harris}, {Hodge}, {de Jong}, {Karachentseva}, {Mateo}, {Stetson},
  {Tavarez}, {Zaritsky}, {Governato}, \& {Quinn}}]{2009ApJS..183...67D}
{Dalcanton}, J.~J., {Williams}, B.~F., {Seth}, A.~C., {et~al.} 2009, \apjs,
  183, 67

\bibitem[{{Dalcanton} {et~al.}(2012){Dalcanton}, {Williams}, {Lang}, {Lauer},
  {Kalirai}, {Seth}, {Dolphin}, {Rosenfield}, {Weisz}, {Bell}, {Bianchi},
  {Boyer}, {Caldwell}, {Dong}, {Dorman}, {Gilbert}, {Girardi}, {Gogarten},
  {Gordon}, {Guhathakurta}, {Hodge}, {Holtzman}, {Johnson}, {Larsen}, {Lewis},
  {Melbourne}, {Olsen}, {Rix}, {Rosema}, {Saha}, {Sarajedini}, {Skillman}, \&
  {Stanek}}]{2012ApJS..200...18D}
{Dalcanton}, J.~J., {Williams}, B.~F., {Lang}, D., {et~al.} 2012, \apjs, 200,
  18

\bibitem[{{Danieli} {et~al.}(2020){Danieli}, {van Dokkum}, {Abraham}, {Conroy},
  {Dolphin}, \& {Romanowsky}}]{2020ApJ...895L...4D}
{Danieli}, S., {van Dokkum}, P., {Abraham}, R., {et~al.} 2020, \apjl, 895, L4

\bibitem[{{de Vaucouleurs} {et~al.}(1991){de Vaucouleurs}, {de Vaucouleurs},
  {Corwin}, {Buta}, {Paturel}, \& {Fouque}}]{1991rc3..book.....D}
{de Vaucouleurs}, G., {de Vaucouleurs}, A., {Corwin}, Herold~G., J., {et~al.}
  1991, {Third Reference Catalogue of Bright Galaxies}

\bibitem[{{Deustua} \& {Mack}(2018)}]{2018wfc..rept....2D}
{Deustua}, S.~E., \& {Mack}, J. 2018, {Comparing the ACS/WFC and WFC3/UVIS
  Calibration and Photometry}, Space Telescope WFC Instrument Science Report, ,

\bibitem[{{Dhawan} {et~al.}(2020){Dhawan}, {Brout}, {Scolnic}, {Goobar},
  {Riess}, \& {Miranda}}]{2020ApJ...894...54D}
{Dhawan}, S., {Brout}, D., {Scolnic}, D., {et~al.} 2020, \apj, 894, 54

\bibitem[{{Di Valentino} {et~al.}(2021){Di Valentino}, {Mena}, {Pan},
  {Visinelli}, {Yang}, {Melchiorri}, {Mota}, {Riess}, \&
  {Silk}}]{2021arXiv210301183D}
{Di Valentino}, E., {Mena}, O., {Pan}, S., {et~al.} 2021, arXiv e-prints,
  arXiv:2103.01183

\bibitem[{{Dolphin}(2016)}]{2016ascl.soft08013D}
{Dolphin}, A. 2016, {DOLPHOT: Stellar photometry}, , , ascl:1608.013

\bibitem[{{Dolphin}(2000)}]{2000PASP..112.1383D}
{Dolphin}, A.~E. 2000, \pasp, 112, 1383

\bibitem[{{Freedman}(2021)}]{2021arXiv210615656F}
{Freedman}, W.~L. 2021, arXiv e-prints, arXiv:2106.15656

\bibitem[{{Freedman} {et~al.}(2019){Freedman}, {Madore}, {Hatt}, {Hoyt},
  {Jang}, {Beaton}, {Burns}, {Lee}, {Monson}, {Neeley}, {Phillips}, {Rich}, \&
  {Seibert}}]{2019ApJ...882...34F}
{Freedman}, W.~L., {Madore}, B.~F., {Hatt}, D., {et~al.} 2019, \apj, 882, 34

\bibitem[{{Freedman} {et~al.}(2020){Freedman}, {Madore}, {Hoyt}, {Jang},
  {Beaton}, {Lee}, {Monson}, {Neeley}, \& {Rich}}]{2020ApJ...891...57F}
{Freedman}, W.~L., {Madore}, B.~F., {Hoyt}, T., {et~al.} 2020, \apj, 891, 57

\bibitem[{{Hamuy} {et~al.}(2006){Hamuy}, {Folatelli}, {Morrell}, {Phillips},
  {Suntzeff}, {Persson}, {Roth}, {Gonzalez}, {Krzeminski}, {Contreras},
  {Freedman}, {Murphy}, {Madore}, {Wyatt}, {Maza}, {Filippenko}, {Li}, \&
  {Pinto}}]{2006PASP..118....2H}
{Hamuy}, M., {Folatelli}, G., {Morrell}, N.~I., {et~al.} 2006, \pasp, 118, 2

\bibitem[{{Hargis} {et~al.}(2020){Hargis}, {Albers}, {Crnojevi{\'c}}, {Sand},
  {Weisz}, {Carlin}, {Spekkens}, {Willman}, {Peter}, {Grillmair}, \&
  {Dolphin}}]{2020ApJ...888...31H}
{Hargis}, J.~R., {Albers}, S., {Crnojevi{\'c}}, D., {et~al.} 2020, \apj, 888,
  31

\bibitem[{{Hatt} {et~al.}(2018{\natexlab{a}}){Hatt}, {Freedman}, {Madore},
  {Beaton}, {Hoyt}, {Jang}, {Lee}, {Monson}, {Rich}, {Scowcroft}, \&
  {Seibert}}]{2018ApJ...861..104H}
{Hatt}, D., {Freedman}, W.~L., {Madore}, B.~F., {et~al.} 2018{\natexlab{a}},
  \apj, 861, 104

\bibitem[{{Hatt} {et~al.}(2018{\natexlab{b}}){Hatt}, {Freedman}, {Madore},
  {Jang}, {Beaton}, {Hoyt}, {Lee}, {Monson}, {Rich}, {Scowcroft}, \&
  {Seibert}}]{2018ApJ...866..145H}
---. 2018{\natexlab{b}}, \apj, 866, 145

\bibitem[{{Hoffmann} \& {Avila}(2017)}]{2017acs..rept....2H}
{Hoffmann}, S.~L., \& {Avila}, R.~J. 2017, {Updated MDRIZTAB Parameters for
  ACS/WFC}, Instrument Science Report ACS 2017-2, ,

\bibitem[{{Hoyt}(2021)}]{2021arXiv210613337H}
{Hoyt}, T.~J. 2021, arXiv e-prints, arXiv:2106.13337

\bibitem[{{Hoyt} {et~al.}(2019){Hoyt}, {Freedman}, {Madore}, {Hatt}, {Beaton},
  {Jang}, {Lee}, {Monson}, {Neeley}, {Rich}, \& {Mager}}]{2019ApJ...882..150H}
{Hoyt}, T.~J., {Freedman}, W.~L., {Madore}, B.~F., {et~al.} 2019, \apj, 882,
  150

\bibitem[{{Hoyt} {et~al.}(2021){Hoyt}, {Beaton}, {Freedman}, {Jang}, {Lee},
  {Madore}, {Monson}, {Neeley}, {Rich}, \& {Seibert}}]{2021arXiv210112232H}
{Hoyt}, T.~J., {Beaton}, R.~L., {Freedman}, W.~L., {et~al.} 2021, arXiv
  e-prints, arXiv:2101.12232

\bibitem[{{Jacobs} {et~al.}(2009){Jacobs}, {Rizzi}, {Tully}, {Shaya},
  {Makarov}, \& {Makarova}}]{2009AJ....138..332J}
{Jacobs}, B.~A., {Rizzi}, L., {Tully}, R.~B., {et~al.} 2009, \aj, 138, 332

\bibitem[{{Jacobs} {et~al.}(2011){Jacobs}, {Tully}, {Rizzi}, {Karachentsev},
  {Chiboucas}, \& {Held}}]{2011AJ....141..106J}
{Jacobs}, B.~A., {Tully}, R.~B., {Rizzi}, L., {et~al.} 2011, \aj, 141, 106

\bibitem[{{Jang} \& {Lee}(2015)}]{2015ApJ...807..133J}
{Jang}, I.~S., \& {Lee}, M.~G. 2015, \apj, 807, 133

\bibitem[{{Jang} \& {Lee}(2017{\natexlab{a}})}]{2017ApJ...836...74J}
---. 2017{\natexlab{a}}, \apj, 836, 74

\bibitem[{{Jang} \& {Lee}(2017{\natexlab{b}})}]{2017ApJ...835...28J}
---. 2017{\natexlab{b}}, \apj, 835, 28

\bibitem[{{Jang} {et~al.}(2018){Jang}, {Hatt}, {Beaton}, {Lee}, {Freedman},
  {Madore}, {Hoyt}, {Monson}, {Rich}, {Scowcroft}, \&
  {Seibert}}]{2018ApJ...852...60J}
{Jang}, I.~S., {Hatt}, D., {Beaton}, R.~L., {et~al.} 2018, \apj, 852, 60

\bibitem[{{Jang} {et~al.}(2021){Jang}, {Hoyt}, {Beaton}, {Freedman}, {Madore},
  {Lee}, {Neeley}, {Monson}, {Rich}, \& {Seibert}}]{2021ApJ...906..125J}
{Jang}, I.~S., {Hoyt}, T.~J., {Beaton}, R.~L., {et~al.} 2021, \apj, 906, 125

\bibitem[{{Karachentsev} {et~al.}(2014){Karachentsev}, {Makarova}, {Tully},
  {Wu}, \& {Kniazev}}]{2014MNRAS.443.1281K}
{Karachentsev}, I.~D., {Makarova}, L.~N., {Tully}, R.~B., {Wu}, P.-F., \&
  {Kniazev}, A.~Y. 2014, \mnras, 443, 1281

\bibitem[{{Krisciunas} {et~al.}(2017){Krisciunas}, {Contreras}, {Burns},
  {Phillips}, {Stritzinger}, {Morrell}, {Hamuy}, {Anais}, {Boldt}, {Busta},
  {Campillay}, {Castell{\'o}n}, {Folatelli}, {Freedman}, {Gonz{\'a}lez},
  {Hsiao}, {Krzeminski}, {Persson}, {Roth}, {Salgado}, {Ser{\'o}n}, {Suntzeff},
  {Torres}, {Filippenko}, {Li}, {Madore}, {DePoy}, {Marshall}, {Rheault}, \&
  {Villanueva}}]{2017AJ....154..211K}
{Krisciunas}, K., {Contreras}, C., {Burns}, C.~R., {et~al.} 2017, \aj, 154, 211

\bibitem[{{Lee} {et~al.}(1993){Lee}, {Freedman}, \&
  {Madore}}]{1993ApJ...417..553L}
{Lee}, M.~G., {Freedman}, W.~L., \& {Madore}, B.~F. 1993, \apj, 417, 553

\bibitem[{{Macri} {et~al.}(2006){Macri}, {Stanek}, {Bersier}, {Greenhill}, \&
  {Reid}}]{2006ApJ...652.1133M}
{Macri}, L.~M., {Stanek}, K.~Z., {Bersier}, D., {Greenhill}, L.~J., \& {Reid},
  M.~J. 2006, \apj, 652, 1133

\bibitem[{{Madore} {et~al.}(2009){Madore}, {Mager}, \&
  {Freedman}}]{2009ApJ...690..389M}
{Madore}, B.~F., {Mager}, V., \& {Freedman}, W.~L. 2009, \apj, 690, 389

\bibitem[{{Mager} {et~al.}(2008){Mager}, {Madore}, \&
  {Freedman}}]{2008ApJ...689..721M}
{Mager}, V.~A., {Madore}, B.~F., \& {Freedman}, W.~L. 2008, \apj, 689, 721

\bibitem[{{Makarov} {et~al.}(2006){Makarov}, {Makarova}, {Rizzi}, {Tully},
  {Dolphin}, {Sakai}, \& {Shaya}}]{2006AJ....132.2729M}
{Makarov}, D., {Makarova}, L., {Rizzi}, L., {et~al.} 2006, \aj, 132, 2729

\bibitem[{{Makarova} {et~al.}(2018){Makarova}, {Makarov}, {Antipova},
  {Karachentsev}, \& {Tully}}]{2018MNRAS.474.3221M}
{Makarova}, L.~N., {Makarov}, D.~I., {Antipova}, A.~V., {Karachentsev}, I.~D.,
  \& {Tully}, R.~B. 2018, \mnras, 474, 3221

\bibitem[{{McQuinn} {et~al.}(2017){McQuinn}, {Skillman}, {Dolphin}, {Berg}, \&
  {Kennicutt}}]{2017AJ....154...51M}
{McQuinn}, K. B.~W., {Skillman}, E.~D., {Dolphin}, A.~E., {Berg}, D., \&
  {Kennicutt}, R. 2017, \aj, 154, 51

\bibitem[{{M{\'e}ndez} {et~al.}(2002){M{\'e}ndez}, {Davis}, {Moustakas},
  {Newman}, {Madore}, \& {Freedman}}]{2002AJ....124..213M}
{M{\'e}ndez}, B., {Davis}, M., {Moustakas}, J., {et~al.} 2002, \aj, 124, 213

\bibitem[{{Monelli} {et~al.}(2010){Monelli}, {Hidalgo}, {Stetson}, {Aparicio},
  {Gallart}, {Dolphin}, {Cole}, {Weisz}, {Skillman}, {Bernard}, {Mayer},
  {Navarro}, {Cassisi}, {Drozdovsky}, \& {Tolstoy}}]{2010ApJ...720.1225M}
{Monelli}, M., {Hidalgo}, S.~L., {Stetson}, P.~B., {et~al.} 2010, \apj, 720,
  1225

\bibitem[{{Planck Collaboration} {et~al.}(2020){Planck Collaboration},
  {Aghanim}, {Akrami}, {Ashdown}, {Aumont}, {Baccigalupi}, {Ballardini},
  {Banday}, {Barreiro}, {Bartolo}, {Basak}, {Battye}, {Benabed}, {Bernard},
  {Bersanelli}, {Bielewicz}, {Bock}, {Bond}, {Borrill}, {Bouchet}, {Boulanger},
  {Bucher}, {Burigana}, {Butler}, {Calabrese}, {Cardoso}, {Carron},
  {Challinor}, {Chiang}, {Chluba}, {Colombo}, {Combet}, {Contreras}, {Crill},
  {Cuttaia}, {de Bernardis}, {de Zotti}, {Delabrouille}, {Delouis}, {Di
  Valentino}, {Diego}, {Dor{\'e}}, {Douspis}, {Ducout}, {Dupac}, {Dusini},
  {Efstathiou}, {Elsner}, {En{\ss}lin}, {Eriksen}, {Fantaye}, {Farhang},
  {Fergusson}, {Fernandez-Cobos}, {Finelli}, {Forastieri}, {Frailis},
  {Fraisse}, {Franceschi}, {Frolov}, {Galeotta}, {Galli}, {Ganga},
  {G{\'e}nova-Santos}, {Gerbino}, {Ghosh}, {Gonz{\'a}lez-Nuevo}, {G{\'o}rski},
  {Gratton}, {Gruppuso}, {Gudmundsson}, {Hamann}, {Handley}, {Hansen},
  {Herranz}, {Hildebrandt}, {Hivon}, {Huang}, {Jaffe}, {Jones}, {Karakci},
  {Keih{\"a}nen}, {Keskitalo}, {Kiiveri}, {Kim}, {Kisner}, {Knox},
  {Krachmalnicoff}, {Kunz}, {Kurki-Suonio}, {Lagache}, {Lamarre}, {Lasenby},
  {Lattanzi}, {Lawrence}, {Le Jeune}, {Lemos}, {Lesgourgues}, {Levrier},
  {Lewis}, {Liguori}, {Lilje}, {Lilley}, {Lindholm}, {L{\'o}pez-Caniego},
  {Lubin}, {Ma}, {Mac{\'\i}as-P{\'e}rez}, {Maggio}, {Maino}, {Mandolesi},
  {Mangilli}, {Marcos-Caballero}, {Maris}, {Martin}, {Martinelli},
  {Mart{\'\i}nez-Gonz{\'a}lez}, {Matarrese}, {Mauri}, {McEwen}, {Meinhold},
  {Melchiorri}, {Mennella}, {Migliaccio}, {Millea}, {Mitra},
  {Miville-Desch{\^e}nes}, {Molinari}, {Montier}, {Morgante}, {Moss}, {Natoli},
  {N{\o}rgaard-Nielsen}, {Pagano}, {Paoletti}, {Partridge}, {Patanchon},
  {Peiris}, {Perrotta}, {Pettorino}, {Piacentini}, {Polastri}, {Polenta},
  {Puget}, {Rachen}, {Reinecke}, {Remazeilles}, {Renzi}, {Rocha}, {Rosset},
  {Roudier}, {Rubi{\~n}o-Mart{\'\i}n}, {Ruiz-Granados}, {Salvati}, {Sandri},
  {Savelainen}, {Scott}, {Shellard}, {Sirignano}, {Sirri}, {Spencer},
  {Sunyaev}, {Suur-Uski}, {Tauber}, {Tavagnacco}, {Tenti}, {Toffolatti},
  {Tomasi}, {Trombetti}, {Valenziano}, {Valiviita}, {Van Tent}, {Vibert},
  {Vielva}, {Villa}, {Vittorio}, {Wandelt}, {Wehus}, {White}, {White},
  {Zacchei}, \& {Zonca}}]{2020A&A...641A...6P}
{Planck Collaboration}, {Aghanim}, N., {Akrami}, Y., {et~al.} 2020, \aap, 641,
  A6

\bibitem[{{Reid} {et~al.}(2019){Reid}, {Pesce}, \&
  {Riess}}]{2019ApJ...886L..27R}
{Reid}, M.~J., {Pesce}, D.~W., \& {Riess}, A.~G. 2019, \apjl, 886, L27

\bibitem[{{Riess} {et~al.}(2016){Riess}, {Macri}, {Hoffmann}, {Scolnic},
  {Casertano}, {Filippenko}, {Tucker}, {Reid}, {Jones}, {Silverman},
  {Chornock}, {Challis}, {Yuan}, {Brown}, \& {Foley}}]{2016ApJ...826...56R}
{Riess}, A.~G., {Macri}, L.~M., {Hoffmann}, S.~L., {et~al.} 2016, \apj, 826, 56

\bibitem[{{Riess} {et~al.}(2021){Riess}, {Yuan}, {Macri}, {Scolnic}, {Brout},
  {Casertano}, {Jones}, {Murakami}, {Breuval}, {Brink}, {Filippenko},
  {Hoffmann}, {Jha}, {Kenworthy}, {Mackenty}, {Stahl}, \&
  {Zheng}}]{2021arXiv211204510R}
{Riess}, A.~G., {Yuan}, W., {Macri}, L.~M., {et~al.} 2021, arXiv e-prints,
  arXiv:2112.04510

\bibitem[{{Rizzi} {et~al.}(2007){Rizzi}, {Tully}, {Makarov}, {Makarova},
  {Dolphin}, {Sakai}, \& {Shaya}}]{2007ApJ...661..815R}
{Rizzi}, L., {Tully}, R.~B., {Makarov}, D., {et~al.} 2007, \apj, 661, 815

\bibitem[{{Rizzi} {et~al.}(2017){Rizzi}, {Tully}, {Shaya}, {Kourkchi}, \&
  {Karachentsev}}]{2017ApJ...835...78R}
{Rizzi}, L., {Tully}, R.~B., {Shaya}, E.~J., {Kourkchi}, E., \& {Karachentsev},
  I.~D. 2017, \apj, 835, 78

\bibitem[{{Salaris} \& {Cassisi}(2005)}]{2005essp.book.....S}
{Salaris}, M., \& {Cassisi}, S. 2005, {Evolution of Stars and Stellar
  Populations}

\bibitem[{{Saviane} {et~al.}(2008){Saviane}, {Momany}, {Da Costa}, {Rich}, \&
  {Hibbard}}]{2008ApJ...678..179S}
{Saviane}, I., {Momany}, Y., {Da Costa}, G.~S., {Rich}, R.~M., \& {Hibbard},
  J.~E. 2008, \apj, 678, 179

\bibitem[{{Schlafly} \& {Finkbeiner}(2011)}]{2011ApJ...737..103S}
{Schlafly}, E.~F., \& {Finkbeiner}, D.~P. 2011, \apj, 737, 103

\bibitem[{{Schweizer} {et~al.}(2008){Schweizer}, {Burns}, {Madore}, {Mager},
  {Phillips}, {Freedman}, {Boldt}, {Contreras}, {Folatelli}, {Gonz{\'a}lez},
  {Hamuy}, {Krzeminski}, {Morrell}, {Persson}, {Roth}, \&
  {Stritzinger}}]{2008AJ....136.1482S}
{Schweizer}, F., {Burns}, C.~R., {Madore}, B.~F., {et~al.} 2008, \aj, 136, 1482

\bibitem[{{Scolnic} {et~al.}(2018){Scolnic}, {Jones}, {Rest}, {Pan},
  {Chornock}, {Foley}, {Huber}, {Kessler}, {Narayan}, {Riess}, {Rodney},
  {Berger}, {Brout}, {Challis}, {Drout}, {Finkbeiner}, {Lunnan}, {Kirshner},
  {Sanders}, {Schlafly}, {Smartt}, {Stubbs}, {Tonry}, {Wood-Vasey}, {Foley},
  {Hand}, {Johnson}, {Burgett}, {Chambers}, {Draper}, {Hodapp}, {Kaiser},
  {Kudritzki}, {Magnier}, {Metcalfe}, {Bresolin}, {Gall}, {Kotak}, {McCrum}, \&
  {Smith}}]{2018ApJ...859..101S}
{Scolnic}, D.~M., {Jones}, D.~O., {Rest}, A., {et~al.} 2018, \apj, 859, 101

\bibitem[{{Serenelli} {et~al.}(2017){Serenelli}, {Weiss}, {Cassisi}, {Salaris},
  \& {Pietrinferni}}]{2017A&A...606A..33S}
{Serenelli}, A., {Weiss}, A., {Cassisi}, S., {Salaris}, M., \& {Pietrinferni},
  A. 2017, \aap, 606, A33

\bibitem[{{Shen} {et~al.}(2021){Shen}, {Danieli}, {van Dokkum}, {Abraham},
  {Brodie}, {Conroy}, {Dolphin}, {Romanowsky}, {Diederik Kruijssen}, \& {Dutta
  Chowdhury}}]{2021ApJ...914L..12S}
{Shen}, Z., {Danieli}, S., {van Dokkum}, P., {et~al.} 2021, \apjl, 914, L12

\bibitem[{{Stetson}(1987)}]{1987PASP...99..191S}
{Stetson}, P.~B. 1987, \pasp, 99, 191

\bibitem[{{Tully} {et~al.}(2016){Tully}, {Courtois}, \&
  {Sorce}}]{2016AJ....152...50T}
{Tully}, R.~B., {Courtois}, H.~M., \& {Sorce}, J.~G. 2016, \aj, 152, 50

\bibitem[{{Tully} {et~al.}(2009){Tully}, {Rizzi}, {Shaya}, {Courtois},
  {Makarov}, \& {Jacobs}}]{2009AJ....138..323T}
{Tully}, R.~B., {Rizzi}, L., {Shaya}, E.~J., {et~al.} 2009, \aj, 138, 323

\bibitem[{{Tully} {et~al.}(2008){Tully}, {Shaya}, {Karachentsev}, {Courtois},
  {Kocevski}, {Rizzi}, \& {Peel}}]{2008ApJ...676..184T}
{Tully}, R.~B., {Shaya}, E.~J., {Karachentsev}, I.~D., {et~al.} 2008, \apj,
  676, 184

\bibitem[{{Tully} {et~al.}(2013){Tully}, {Courtois}, {Dolphin}, {Fisher},
  {H{\'e}raudeau}, {Jacobs}, {Karachentsev}, {Makarov}, {Makarova},
  {Mitronova}, {Rizzi}, {Shaya}, {Sorce}, \& {Wu}}]{2013AJ....146...86T}
{Tully}, R.~B., {Courtois}, H.~M., {Dolphin}, A.~E., {et~al.} 2013, \aj, 146,
  86

\bibitem[{{Williams} {et~al.}(2014){Williams}, {Lang}, {Dalcanton}, {Dolphin},
  {Weisz}, {Bell}, {Bianchi}, {Byler}, {Gilbert}, {Girardi}, {Gordon},
  {Gregersen}, {Johnson}, {Kalirai}, {Lauer}, {Monachesi}, {Rosenfield},
  {Seth}, \& {Skillman}}]{2014ApJS..215....9W}
{Williams}, B.~F., {Lang}, D., {Dalcanton}, J.~J., {et~al.} 2014, \apjs, 215, 9

\bibitem[{{Wu} {et~al.}(2014){Wu}, {Tully}, {Rizzi}, {Dolphin}, {Jacobs}, \&
  {Karachentsev}}]{2014AJ....148....7W}
{Wu}, P.-F., {Tully}, R.~B., {Rizzi}, L., {et~al.} 2014, \aj, 148, 7

\bibitem[{{York} {et~al.}(2000){York}, {Adelman}, {Anderson}, {Anderson},
  {Annis}, {Bahcall}, {Bakken}, {Barkhouser}, {Bastian}, {Berman}, {Boroski},
  {Bracker}, {Briegel}, {Briggs}, {Brinkmann}, {Brunner}, {Burles}, {Carey},
  {Carr}, {Castander}, {Chen}, {Colestock}, {Connolly}, {Crocker}, {Csabai},
  {Czarapata}, {Davis}, {Doi}, {Dombeck}, {Eisenstein}, {Ellman}, {Elms},
  {Evans}, {Fan}, {Federwitz}, {Fiscelli}, {Friedman}, {Frieman}, {Fukugita},
  {Gillespie}, {Gunn}, {Gurbani}, {de Haas}, {Haldeman}, {Harris}, {Hayes},
  {Heckman}, {Hennessy}, {Hindsley}, {Holm}, {Holmgren}, {Huang}, {Hull},
  {Husby}, {Ichikawa}, {Ichikawa}, {Ivezi{\'c}}, {Kent}, {Kim}, {Kinney},
  {Klaene}, {Kleinman}, {Kleinman}, {Knapp}, {Korienek}, {Kron}, {Kunszt},
  {Lamb}, {Lee}, {Leger}, {Limmongkol}, {Lindenmeyer}, {Long}, {Loomis},
  {Loveday}, {Lucinio}, {Lupton}, {MacKinnon}, {Mannery}, {Mantsch}, {Margon},
  {McGehee}, {McKay}, {Meiksin}, {Merelli}, {Monet}, {Munn}, {Narayanan},
  {Nash}, {Neilsen}, {Neswold}, {Newberg}, {Nichol}, {Nicinski}, {Nonino},
  {Okada}, {Okamura}, {Ostriker}, {Owen}, {Pauls}, {Peoples}, {Peterson},
  {Petravick}, {Pier}, {Pope}, {Pordes}, {Prosapio}, {Rechenmacher}, {Quinn},
  {Richards}, {Richmond}, {Rivetta}, {Rockosi}, {Ruthmansdorfer}, {Sand ford},
  {Schlegel}, {Schneider}, {Sekiguchi}, {Sergey}, {Shimasaku}, {Siegmund},
  {Smee}, {Smith}, {Snedden}, {Stone}, {Stoughton}, {Strauss}, {Stubbs},
  {SubbaRao}, {Szalay}, {Szapudi}, {Szokoly}, {Thakar}, {Tremonti}, {Tucker},
  {Uomoto}, {Vanden Berk}, {Vogeley}, {Waddell}, {Wang}, {Watanabe},
  {Weinberg}, {Yanny}, {Yasuda}, \& {SDSS Collaboration}}]{2000AJ....120.1579Y}
{York}, D.~G., {Adelman}, J., {Anderson}, John~E., J., {et~al.} 2000, \aj, 120,
  1579

\end{thebibliography}
\bibliographystyle{aasjournal}

\end{document}